\documentclass[draft]{agujournal2019}
\draftfalse

\usepackage{apacite}
\usepackage{url} 
\usepackage{graphics}

\usepackage{amsfonts,amssymb}
\usepackage{amsmath}
\usepackage{float}
\usepackage{bm}
\usepackage{algorithm}
\usepackage[noend]{algpseudocode}
\usepackage{color}
\usepackage{booktabs, caption, makecell}
\usepackage{threeparttable}
\usepackage{multirow}
\usepackage{multicol}

\journalname{Water Resources Research}

\begin{document}

\justify

\title{Integration of adversarial autoencoders with residual dense convolutional networks for estimation of non-Gaussian hydraulic conductivities}

\authors{Shaoxing Mo\affil{1,2}, Nicholas Zabaras\affil{2}, Xiaoqing Shi\affil{1}, and Jichun Wu\affil{1}}

\affiliation{1}{Key Laboratory of Surficial Geochemistry of Ministry of Education, School of Earth Sciences and Engineering, Nanjing University, Nanjing, China.}

\affiliation{2}{Center for Informatics and Computational Science, University of Notre Dame, Notre Dame, IN, USA.}
 
\correspondingauthor{Nicholas Zabaras}{nzabaras@gmail.com}
\correspondingauthor{Xiaoqing Shi}{shixq@nju.edu.cn}

\begin{keypoints}
    \item A convolutional adversarial autoencoder is developed to parameterize non-Gaussian conductivity fields with multimodal distributions
    \item A deep residual dense convolutional network is introduced as a surrogate of the forward physics-based model
    \item The integrated method is tested with inverse problems for the estimation of non-Gaussian conductivities in solute transport modeling
\end{keypoints}

\begin{abstract}
Inverse modeling for the estimation of non-Gaussian hydraulic conductivity fields in subsurface flow and solute transport models remains a challenging problem.
This is mainly due to the non-Gaussian property, the non-linear physics, and the fact that many repeated evaluations of the forward model are often required. In this study, we develop a convolutional adversarial autoencoder (CAAE) to parameterize  non-Gaussian conductivity fields with heterogeneous conductivity within each facies using a low-dimensional latent representation. In addition, a deep residual dense convolutional network (DRDCN) is proposed for surrogate modeling of forward models with high-dimensional and highly-complex mappings. The two networks are both based on a multilevel residual learning architecture called residual-in-residual dense block. The multilevel residual learning strategy and the dense connection structure ease the training of deep networks, enabling us to efficiently build deeper networks that have an essentially increased capacity for approximating mappings of very high-complexity. The CCAE and DRDCN networks are incorporated into an iterative ensemble smoother to formulate an inversion framework. 
The numerical experiments performed using 2-D and 3-D solute transport models  illustrate the performance of the integrated method.
The obtained results indicate that the CAAE is a robust parameterization method for non-Gaussian conductivity fields with different heterogeneity patterns. The DRDCN is able to obtain accurate approximations of the forward models with high-dimensional and highly-complex mappings using relatively limited training data. The CAAE and DRDCN methods together significantly reduce the computation time required to achieve accurate inversion results.

\end{abstract}

\section{Introduction}

Groundwater flow and solute transport models are used widely to help understand subsurface processes and make science-informed decisions for groundwater resource management. 
Reliable model predictions that well reproduce the phenomena of interest require a good characterization of the hydraulic conductivity field as it greatly 
influences groundwater flow and solute transport. 
In many practical cases, such as aquifers in fluvial deposits where several highly contrasting facies coexist, it may be unrealistic to model the log-conductivity as a multivariate Gaussian. It has been shown that a multimodal 
distribution can better characterize the connectivity between different facies and the multimodality in highly heterogeneous conductivity fields~\cite{GOMEZHERNANDEZ199847,Journel1993,KERROU2008147,ZHOU201422}. 

In this study, we are concerned with an inverse estimation of non-Gaussian conductivity fields with heterogeneous conductivity within each facies using  hydraulic head and solute concentration data. This is in contrast to 
categorical fields with homogeneous conductivity within each facies (e.g., the binary field)~\cite{Winter2003}, although they both have multimodal distributions. In practice, the conductivity field is estimated via inverse modeling. Commonly used inverse methods are, for example, Markov chain Monte Carlo methods~\cite{VRUGT2016273} and the ensemble-based data assimilation methods such as ensemble smoother~\cite{VanLeeuwen1996}, ensemble Kalman filter~\cite{Evensen1994}, and their variants~\cite{Emerick2013,laloy2013,sun2009,xu2018,zhang2018ILUES,zhou2011}.
Considering the strong conductivity heterogeneity, the inverse problem 
is usually high-dimensional. Thus, it often requires a large number of forward model runs to obtain converged inversion results. 

To relieve the large computational cost, parameterization 
methods are commonly used together with surrogate models within the inversion framework. A parameterization method aims 
to represent the spatially correlated property field using a low-dimensional latent vector~\cite{LINDE201586,Oliver2011,ZHOU201422}. 
It can also 1) mitigate the potential ill-posedness of the inverse problem, 2) ensure that the updated fields in the 
inversion process satisfy the prior distribution assumptions (e.g., non-Gaussian) imposed on the unknown field, and 3) produce normally distributed latent variables, allowing direct use of second-order inverse methods (e.g., ensemble smoother and ensemble Kalman filter), although these may be at the cost of increased nonlinearity for the inverse problems~\cite{LALOY2019}. In previous studies involving the estimation of non-Gaussian conductivity fields, as noted in the review by~\citeA{LINDE201586}, one popular solution is to utilize multiple-point statistics (MPS) simulation in the inversion to generate conductivity field realizations that honor the  non-Gaussian prior determined by a training image~\cite{Hansen2012,LALOY201657,Mariethoz2010,Zahner2016}. While such MPS-based inverse methods can generally provide reliable estimations, they may be computationally intensive~\cite{laloy2018}. 
A surrogate method aims to replace the computationally expensive forward model with an accurate but cheap-to-run 
surrogate model during the inversion~\cite{Asher2015,Razavi2012}. Although such combinations 
of methods for inverse modeling have been intensively studied 
for problems with Gaussian conductivity fields~\cite{Chang2017,Elsheikh2014,ju2018,laloy2013,zhang2015,zhang2016}, previous 
studies on problems with non-Gaussian conductivity fields often relied on the inverse methods solely without using the 
parameterization and surrogate methods. The development of parameterization and surrogate methods for such non-Gaussian problems remains an open problem due to two major challenges.

First, most existing parameterization methods fail to work for non-Gaussian conductivity fields. Previous works on parameterizing the conductivity fields in inversion has relied on, for example, principal component analysis and its variants~\cite{MA20117311,Sarma2008,Vo2014,zhang2004}. While these methods are well suited for Gaussian random fields, their performance for complex non-Gaussian fields deserves further improvement~\cite{Canchumuni201987,Chan2017,laloy2017,Liu2019}. 
Inspired by the recent success of deep learning in various areas including Earth science~\cite{Bergeneaau0323,Reichstein2019,ZUO20191} and hydrology~\cite{shen2018}, its application in 
parameterization of non-Gaussian conductivity fields has been reported in many recent studies~\cite{Canchumuni2019941,Canchumuni201987,Chan2017,Chan2018,Chan2019,laloy2017,laloy2018,Liu2019}. Among 
these applications, generative adversarial network (GAN)~\cite{Goodfellow2014} and variational antoencoder (VAE)~\cite{Kingma2014} 
are the two most popular network architectures. After training the network, these methods take random realizations of a low-dimensional vector as input and then 
generate new realizations of the conductivity field having similar features with those found in the training data. The quality of 
the generated realizations was shown to be superior to those from traditional parameterization methods~\cite{Canchumuni201987,Chan2017,laloy2017,Liu2019}. 
However, these methods focused on  categorical conductivity fields with homogeneous conductivity within each facies. Their applicability to  non-Gaussian fields with heterogeneous conductivity within each facies, which is more challenging, remains to be further investigated. \citeA{laloy2018} and~\citeA{Liu2019} further tested their methods' potential for such continuous fields. While the results suggested a promising performance of the deep learning-based parameterization methods, no inversion results were presented. \citeA{Canchumuni201987} considered channelized conductivity fields with a bimodal distribution. They used a VAE network to generate binary channelized fields and the uncertain parameters to be estimated in the inversion are the latent variables and the permeability values at all grid points, with the inverse problem remaining high-dimensional.

Second, most existing surrogate methods suffer from the curse of dimensionality~\cite{Asher2015,Razavi2012} and fail to efficiently obtain 
accurate approximations when the input-output relations are highly-nonlinear~\cite{liao2017,lin2009,mo2017}. The curse of dimensionality is caused by the 
exponentially increased computational cost required for accurate surrogate construction as the input dimensionality (i.e., the number of uncertain variables considered) increases. 
Due to the strongly heterogeneous nature of the conductivity field, it is often required to use a large number of stochastic degrees of freedom to accurately represent the heterogeneity. 
The highly-nonlinear outputs here arise because the high-conductivity regions in a non-Gaussian aquifer result in preferential paths for the groundwater flow and solute transport.  The 
two factors together make the commonly used surrogate methods, such as Gaussian processes~\cite{Rasmussen2006} and 
polynomial chaos expansion~\cite{xiu2002}, difficult to work. 
Deep neural networks have already exhibited a promising and impressive performance for surrogate modeling of forward 
models with high-dimensional input and output fields~\cite{Kani2019,mo2019inverse,mo2019UQ,Sun2018,Tripathy2018,Zhong2019,ZHU2018415,zhu2019physics}. 
For example, in~\citeA{Tripathy2018} a deep neural network was proposed to build a surrogate model for a single-phase flow forward model. 
In~\citeA{Sun2018} and~\citeA{Zhong2019}, their surrogate methods for a single-phase flow forward model and a multiphase flow forward model, respectively, 
were based on an adversarial network framework. In our previous studies~\cite{mo2019inverse,mo2019UQ,ZHU2018415,zhu2019physics}, 
a deep dense convolutional network (DDCN), which is based on a dense connection structure~\cite{Huang2016} for better information flow efficiency, was employed as the surrogate modeling framework. It showed a good performance 
in efficiently obtaining accurate surrogates of  various forward models 
with high-dimensional input-output mappings. However, these methods were tested on forward models with Gaussian 
conductivity fields~\cite{Kani2019,mo2019inverse,mo2019UQ,Tripathy2018,ZHU2018415,zhu2019physics} or on a single-phase 
flow model with binary channelized conductivity fields~\cite{zhu2019physics}. As shown in a case study below, the application of the DDCN surrogate method to solute transport modeling with non-Gaussian conductivity fields may lead to large approximation errors.  

In this work, we develop a convolutional adversarial autoencoder (CAAE) to parameterize  non-Gaussian conductivity fields with multimodal distributions. 
We transform a fully-connected adversarial autoencoder~\cite{Makhzani2016} to  a convolutional network so as 
to improve its scalability for larger-size inputs. 
In addition, we propose a deep residual dense convolutional network (DRDCN) for 
efficient surrogate modeling of forward models with high-dimensional and highly-complex mappings. Although deeper networks have the potential to 
substantially improve the network's performance, they can be difficult to train. We adopt in DRDCN a multilevel residual learning structure~\cite{Wang-etal2018}. The residual learning strategy has been shown to be an effective solution to ease the training of very deep networks~\cite{He2016CVPR,He2016ECCV,Simonyan2015,Szegedy_2015_CVPR,Wang-etal2018}. The multilevel residual learning structure is also implemented in the CAAE network. The CAAE and DRDCN networks are 
combined with an iterative local updating ensemble smoother (ILUES) algorithm~\cite{zhang2018ILUES} to formulate 
an efficient CAAE-DRDCN-ILUES inversion framework.  The overall integrated method 
is demonstrated using 2-D  and 3-D solute transport modeling with  non-Gaussian conductivity fields that have different heterogeneity patterns.  

In summary, three major innovative contributions are addressed in this study. First, we develop a CAAE method for parameterization of non-Gaussian conductivity fields with heterogeneous conductivity within each facies that is suitable in the context of inverse modeling. 
Second, we adopt a multilevel residual strategy in our previous DDCN method~\cite{mo2019inverse,mo2019UQ,ZHU2018415,zhu2019physics} to introduce a new DRDCN method with a substantially improved 
performance for surrogate modeling of highly-complex mappings. Finally and most importantly, to the best of our knowledge, we present the first attempt to incorporate simultaneously the parameterization and surrogate methods to perform inversion for non-Gaussian conductivities in solute transport modeling. 

The rest of the paper is organized as follows. In section~\ref{sec:problem-def}, we introduce a solute transport model 
and define the problem of interest. The CAAE-DRDCN-ILUES inversion framework is presented in section~\ref{sec:methods}. 
Then in sections~\ref{sec:application} and~\ref{sec:results&discussion}, the proposed method is evaluated using two synthetic examples. The conclusions are summarized in the last section.

\section{Problem Definition}\label{sec:problem-def}
We consider  solute transport in heterogeneous porous media under a steady-state groundwater flow condition. It is 
assumed that the transport of solute is driven by  advection and dispersion. The governing equations 
for the steady-state flow and solute transport are written as~\cite{zheng1999}
\begin{linenomath*}
\begin{equation}\label{eq:flow}
     \nabla\cdot(K\nabla h)=0,
\end{equation}
\end{linenomath*}
and 
\begin{linenomath*}
\begin{equation}\label{eq:adv-disp}
    \frac{\partial(\phi c)}{\partial t}=\nabla\cdot(\phi \boldsymbol{\alpha}\nabla c)-\nabla\cdot(\phi \bm vc)+r_s,
\end{equation}
\end{linenomath*}
respectively. Here $K$~(LT$^{-1}$) is the hydraulic conductivity, $h$~(L) is the hydraulic head, $\phi$~(-) is the 
effective porosity, $c$~(ML$^{-3}$) is the solute concentration, $t$~(T) denotes time, $r_s$~(ML$^{-3}$T$^{-1}$) 
is the sink/source, and $\boldsymbol{\alpha}$~(L$^2$T$^{-1}$) is the dispersion 
tensor determined by the pore space flow velocity $\bm v$~(LT$^{-1}$), and longitudinal ($\alpha_L$;~L), transverse ($\alpha_T$;~L), and vertical ($\alpha_V$;~L) dispersivities. The two equations 
are coupled through the velocity $\bm v=-\frac{K}{\phi}\nabla h$. The flow and solute transport equations are numerically solved using the MODFLOW~\cite{Harbaugh2000} and MT3DMS~\cite{zheng1999} simulators, respectively. 

We are concerned with an inverse problem of characterizing the heterogeneous conductivity field using measurements of the hydraulic head and concentration.  The underlying conductivity fields of interest are non-Gaussian fields with a multimodal distribution. The inverse modeling is performed using the ILUES inversion algorithm~\cite{zhang2018ILUES} which 
has shown a promising performance for high-dimensional and highly-nonlinear inverse problems~\cite{mo2019inverse,zhang2018ILUES}.

\section{Methodology}\label{sec:methods}

\subsection{ILUES for Inverse Modeling}\label{sec:ILUES}
The ILUES algorithm assimilates the output measurements $\bm{d}\in\mathbb{R}^{N_d}$ for multiple times with 
an inflated covariance matrix of the measurement errors to avoid overweighing the measurements~\cite{zhang2018ILUES}. 
The inflated covariance matrix is often taken as $\tilde{\mathbf{C}}_{\rm{D}}=N_{\rm{iter}}\mathbf{C}_{\rm{D}}$, where  $\mathbf C_{\rm{D}}$  is 
the original covariance matrix of the measurement errors and $N_{\rm{iter}}$ is the number of iterations~\cite{Emerick2013,zhang2018ILUES}. 
To better handle high-dimensional and highly-nonlinear problems, ILUES also adopts a local updating scheme, which updates each 
input sample $\bm{m}\in\mathbb{R}^{N_m}$ (in the CAAE-DRDCN-ILUES framework, $\bm{m}$ refers to the latent variables, see section~\ref{sec:CAAE-DRDCN-ILUES}) in the ensemble locally using its neighboring samples rather than all samples 
in the ensemble. Formally, given an ensemble of $N_e$ input samples $\mathbf M=[\bm m_1,\ldots,\bm m_{N_e}]$, it first 
identifies a local ensemble for each sample $\bm m_i\in\mathbf M$ based on the following metric~\cite{zhang2018ILUES}
\begin{linenomath*}
\begin{equation}\label{eq:J-value}
    J(\bm m)=\frac{J_d(\bm m)}{J_d^{\text{max}}} + \frac{J_m(\bm m)}{J_m^{\text{max}}},
\end{equation}
\end{linenomath*}
where $J_d(\bm m)=[f(\bm m)-\bm d]^{\top}\mathbf C_{\rm{D}}^{-1}[f(\bm m)-\bm d]$ quantifies the mismatch between 
the model responses $f(\bm m)$ and measurements $\bm d$, and $J_m(\bm m)=(\bm m-\bm m_i)^{\top}\mathbf C_{\text{MM}}^{-1}(\bm m-\bm m_i)$ 
is the distance between the sample $\bm m\in\mathbf M$ and $\bm m_i$. Here, $\mathbf C_{\rm{MM}}$ is the autocovariance matrix 
of the input parameters in $\mathbf M$, $J_d^{\rm{max}}$ and $J_m^{\text{max}}$ are the maximum values of $J_d(\cdot)$ and $J_m(\cdot)$, 
respectively. Based on the $J$ values, we select $N_l=\beta_l N_e,\;(\beta_l\in(0,1])$ samples as the local ensemble of $\bm m_i$ 
using a roulette wheel selection operator~\cite{Lipowski2012}, in which the selection probability 
of the $i$th individual is given as $P_i=\rho_i/\sum_{j=1}^{N_e}\rho_j$, $i=1,\ldots,N_e$, where $\rho_j=1/J(\bm m_j)$~\cite{mo2019inverse}. 
A local ensemble factor of $\beta_l=0.1$ suggested in~\citeA{zhang2018ILUES} is used. 
 
 Let superscripts $l$, $f$, and $a$ denote the local ensemble, current and updated samples, respectively. The ILUES first updates the local 
 ensemble of each sample $\bm{m}_i^f\in\mathbf{M}^f$, that is, $\mathbf{M}_i^{l,f}$, by using the usual ensemble smoother scheme~\cite{Emerick2013,zhang2018ILUES}:
\begin{linenomath*}
 \begin{equation}\label{eq:es}
     \bm m_{i,j}^a=\bm m_{i,j}^f+\mathbf{C}_{\text{MD}}^{l,f}\big(\mathbf{C}_{\text{DD}}^{l,f}+\tilde{\mathbf C}_{\rm D}\big)^{-1}\big[\bm{d}_j-f(\bm{m}_{i,j}^f)\big],
 \end{equation}
 \end{linenomath*}
 for $j=1,\ldots,N_l$. Here $\mathbf{C}_{\text{MD}}^{l,f}$ is the cross-covariance matrix between $\mathbf M_i^{l,f}$ 
 and $\mathbf D_i^{l,f}=\big[f(\bm m_{i,1}^f),\ldots,f(\bm m_{i,N_l}^f)\big]$, $\mathbf{C}_{\text{DD}}^{l,f}$ is the 
 autocovariance matrix of $\mathbf D_i^{l,f}$, and $\bm d_j=\bm d+\tilde{\mathbf C}_{\rm D}^{1/2}\bm r_{N_d}$, $\bm r_{N_d}\sim \mathcal N(\bm 0,\mathbf I)$, 
 is the $j$-th realization of the measurements. The update of $\bm m_i^f$, $\bm m_i^a$, is then generated from its 
 updated local ensemble $\mathbf M_i^{l,a}=[\bm m_{i,1}^a,\ldots,\bm m_{i,N_l}^a]$ through a probabilistic scheme~\cite{mo2019inverse}. 
 One update iteration of ILUES is summarized in Algorithm~\ref{algor:ilues-update}. More details regarding ILUES 
 can be found in~\citeA{zhang2018ILUES} and~\citeA{mo2019inverse}.

\begin{algorithm}[htb]
\caption{One update iteration in iterative local updating ensemble smoother. RWS: roulette wheel selection.}
\label{algor:ilues-update}
\begin{algorithmic}[1]

\Require Measurements $\bm{d}$, ensemble size $N_e$, local ensemble size factor $\beta_l$, current input 
ensemble $\mathbf M^f=[\bm m_1^f,\ldots,\bm m_{N_e}^f]$ and output ensemble $\mathbf D^f=\big[f(\bm m_1^f),\ldots,f(\bm m_{N_e}^f)\big]$.

\State $N_l\gets\beta_l N_e$.

  \For{$i=1,\ldots,N_e$} \Comment Update each sample using its local ensemble
     \State Given $\bm m_i^f$, compute the $J$ values for samples in $\mathbf M^f$ using equation~(\ref{eq:J-value}).
     
     \State Choose the local ensemble of $\bm m_i^f$, $\mathbf M_i^{l,f}=[\bm m_{i,1}^f,\ldots,\bm m_{i,N_l}^f]$, using RWS based on the $J$ values.
     
     \State Obtain the updated local ensemble $\mathbf M_i^{l,a}=[\bm m_{i,1}^a,\ldots,\bm m_{i,N_l}^a]$ using equation~(\ref{eq:es}).
     
     \State Randomly draw a sample $\bm m_{i,j}^a\in \mathbf M_i^{l,a}$ and run the forward model $f(\bm m_{i,j}^a)$. \Comment{In CAAE-DRDCN-ILUES $f(\cdot)$ is computed using the surrogate model instead (see section~\ref{sec:CAAE-DRDCN-ILUES})}
     
     \State $r_{\rm{a}}=\text{min}\big\{1,\exp{\big[-0.5\big(J_d(\bm m_{i,j}^a)-J_d(\bm m_i^f)\big)\big]}\big\}.$ \Comment{$J_d(\cdot)$ 
     is defined in equation~(\ref{eq:J-value})}.
     
     \State 
     $\big[\bm m_i^a,f(\bm m_i^a)\big]=
    \begin{cases}
        \big[\bm m_{i,j}^a,f(\bm m_{i,j}^a)\big],& \gamma\leq r_{\rm{a}}\\
       \big[\bm m_i^f,f(\bm m_i^f)\big],& \gamma> r_{\rm{a}}
    \end{cases},$ where $\gamma\sim\mathcal U[0,1]$.
     
  \EndFor
  \State \textbf{end for}
  \State $\mathbf M^a=[\bm m_1^a,\ldots,\bm m_{N_e}^a]$, $\mathbf D^a=\big[f(\bm m_1^a),\ldots,f(\bm m_{N_e}^a)\big]$. 

\State \textbf{return} $\mathbf M^a$, $\mathbf D^a$ \Comment{The updated input and output ensembles}
\end{algorithmic}
\end{algorithm}
 
For high-dimensional inverse problems, large ensemble size and iteration number are usually needed for ILUES 
to obtain converged and reliable inversion results, resulting in a large computational cost in forward model 
runs. To  reduce the computational burden, we propose a CAAE network for parameterizing 
the high-dimensional conductivity field using a low-dimensional latent vector and a DRDCN network 
to build an accurate but fast-to-run substitution of the forward model in the ILUES algorithm. 

\subsection{DRDCN for Surrogate Modeling}\label{sec:DRDCN}
In the surrogate modeling task, we build a surrogate model to approximate the mapping between the input 
conductivity field and the output hydraulic head and concentration fields. In our previous 
studies~\cite{mo2019inverse,mo2019UQ,ZHU2018415,zhu2019physics}, we transformed the surrogate modeling task for problems 
with high-dimensional input and output fields  in a 2-D domain to an image-to-image regression problem by using a 
DDCN network. 
In this network, the input and output fields were treated as images. Denoting $H\times W$ as the 
spatial discretization resolution of the domain, $\mathbf{x}\in\mathbb{R}^{n_x\times H\times W}$ and $\mathbf{y}\in\mathbb{R}^{n_y\times H\times W}$ 
as the input and output fields, respectively. Then the surrogate modeling task for approximating the input-output mapping, 
\begin{linenomath*}
\begin{equation}\label{eq:mapping}
    f:~\mathbb{R}^{n_x\times H\times W}\rightarrow{\mathbb{R}^{n_y\times H\times W}},
\end{equation}
\end{linenomath*}
was transformed to an image regression problem between $n_x$ input images and $n_y$ output images with a 
resolution of $H\times W$, where $n_x$ and $n_y$ are the number of the input and output fields, respectively. 
It is straightforward to generalize to a 3-D domain by adding an extra depth axis to the images, that is, $\mathbb{R}^{n_x\times D\times H\times W}\rightarrow{\mathbb{R}^{n_y\times D\times H\times W}}$. 

In order to further improve the performance of DDCN in problems with highly-complex mappings, we 
adopt a novel basic block called `residual-in-residual dense block' proposed in~\citeA{Wang-etal2018} for 
image super-resolution problems to formulate our DRDCN framework.

\subsubsection{Residual-in-Residual Dense Block}\label{sec:RRDB}
\begin{figure}[h!]
    \centering
    \includegraphics[width=.9\textwidth]{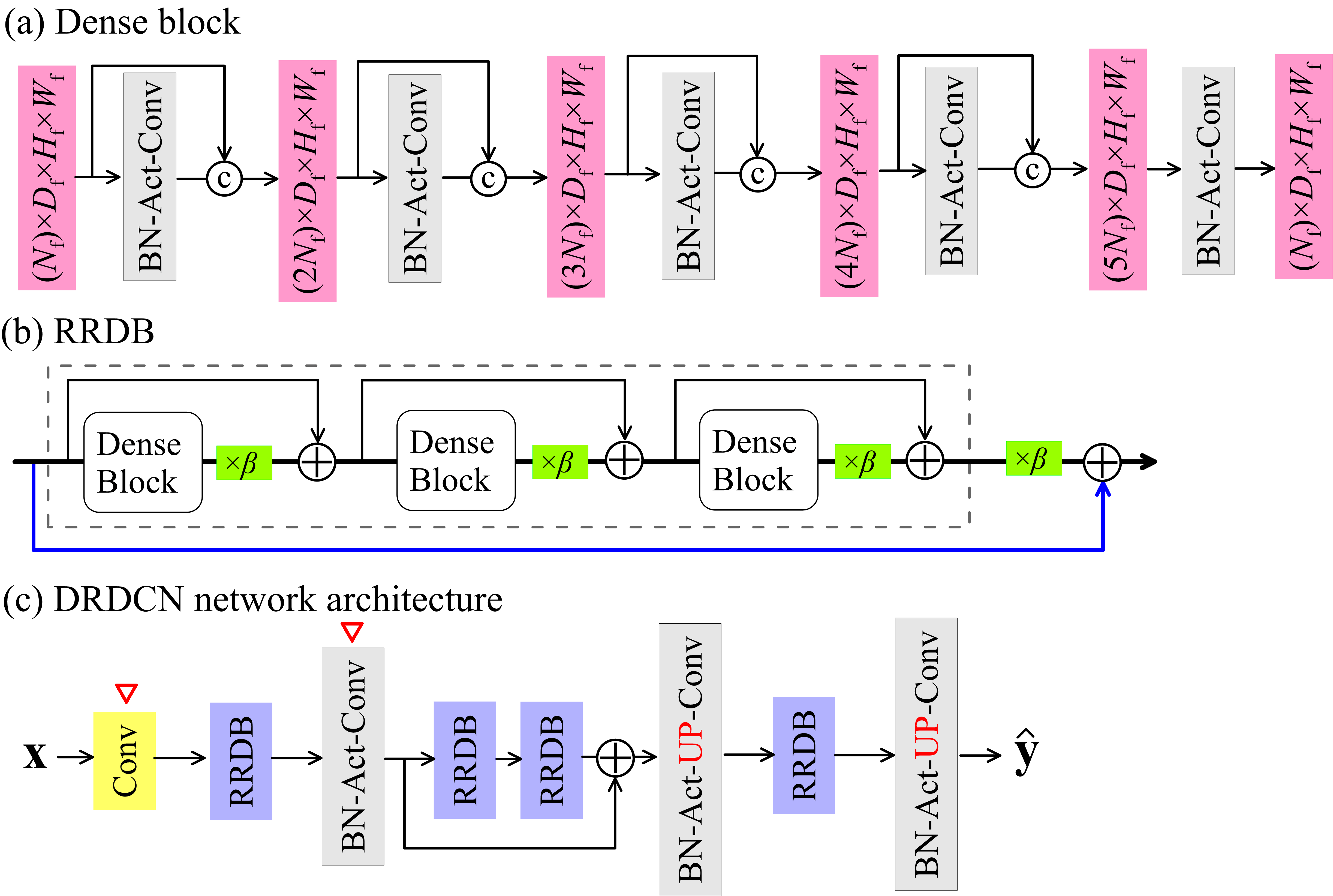}
    \caption{(a) A dense block with five layers. Each layer contains three operations (i.e., BN, Act, and Conv) and outputs $N_{\mathrm{f}}$ feature maps with size $D_{\mathrm{f}}\times H_{\mathrm{f}}\times W_{\mathrm{f}}$, which are concatenated ($\copyright$) 
    with its input feature maps. The concatenated feature maps are treated as the input to the next layer. (b) A residual-in-residual dense block (RRDB) with three residual dense blocks. 
    The output of a dense block is scaled down by multiplying with a factor $\beta\in(0,1]$ before adding ($\bigoplus$) to its input. 
    (c) Architecture of the deep residual dense convolutional network (DRDCN). 
    The feature map size is halved in Conv ($\bigtriangledown$) with stride 2 and doubled by using the nearest upsampling (UP) operation.}
    \label{fig:net4surrogate}
\end{figure}

A dense block introduces connections between non-adjacent layers aiming to fully exploit the hierarchical features 
from the outputs of preceding layers~\cite{Huang2016}. Let $\mathbf{z}^{(i)}$ ($i=1,\ldots,L$) denote the output 
feature maps of the $i$th layer in the dense block, where $L$ is number of layers. $\mathbf{z}^{(i)}$ is obtained by taking the concatenation of the output feature maps 
from its preceding layers as input, as represented by
\begin{linenomath*}
\begin{equation}\label{eq:DB}
    \mathbf{z}^{(i)}=\mathcal{H}([\mathbf{z}^{(0)},\ldots,\mathbf{z}^{(i-1)}]),
\end{equation}
\end{linenomath*}
where $\mathbf{z}^{(0)}$ represents the input to the dense block, and $\mathcal{H}$ denotes operations on the input feature maps, including batch normalization (BN)~\cite{Ioffe2015}, 
followed by nonlinear activation and convolution (Conv)~\cite{Goodfellow-et-al-2016}. The ReLU (Rectified Linear Unit) function defined as $\texttt{ReLU}(x)=\max(0,x)$ is commonly used. Alternatively, a new activation function Mish~\cite{Misra2019arXiv}, which is defined as $\texttt{Mish}(x)=x\cdot\tanh(\ln(1+e^x))$, can be used. It has been shown to perform generally better than ReLU in many deep networks across many datasets~\cite{Misra2019arXiv}. 
A dense block with $L=5$ layers is illustrated in Figure~\ref{fig:net4surrogate}a. 

It has been shown that deeper networks   have the potential to  
better approximate mappings of  high complexity, however,  they can be difficult to train~\cite{Simonyan2015,Szegedy_2015_CVPR,Wang-etal2018}. 
To efficiently train a deeper network, we adopt   a multilevel residual learning structure, that is, the residual-in-residual 
dense block proposed in~\citeA{Wang-etal2018}. In the residual learning framework, it has been shown that the residual mapping is much 
easier to learn than the original mapping~\cite{He2016CVPR,He2016ECCV}. Specifically, let $f(\mathbf{x})$ 
denote the desired underlying mapping to fit. Then the stacked layers (not necessarily the entire network) learn the residual mapping  
 $g(\mathbf{x}):=f(\mathbf{x})-\mathbf{x}$.  The original mapping $f(\mathbf{x})$ is then 
recast as $g(\mathbf{x})+\mathbf{x}$. Such a residual learning strategy can help alleviate the 
gradient vanishing problem for deep network training~\cite{He2016CVPR} and thus ease the training of very 
deep networks to achieve improved accuracy~\cite{He2016CVPR,He2016ECCV,Ledig2017CVPR,Simonyan2015,Wang-etal2018}. 

The architecture of residual-in-residual dense block is shown in Figure~\ref{fig:net4surrogate}b. It consists of a stack of residual dense blocks, 
where the residual learning is used in two levels, resulting in a residual-in-residual structure. 
 That is, the residual learning implemented to the dense block results in a residual dense block; and that implemented to the stacked residual dense blocks results in a residual-in-residual dense block. 
The number of input and output feature maps of a dense block is $N_{\rm f}$ (Figure~\ref{fig:net4surrogate}a) and is set to $N_{\rm f}=48$. In addition to the residual-in-residual structure, a residual scaling technique~\cite{Szegedy2016} is also employed in the
residual-in-residual dense block to further increase the training stability~\cite{Wang-etal2018}. Formally, this is done by scaling down the residual $g(\mathbf{x})$ 
by a factor $\beta\in(0,1]$ before adding to $\mathbf{x}$ (Figure~\ref{fig:net4surrogate}b). A factor of $\beta=0.2$ suggested in~\citeA{Wang-etal2018} 
is used in our network.

\subsubsection{DRDCN Networks Based on Residual-in-Residual Dense Blocks}\label{sec:DRDCN-arc}
We employ the residual-in-residual dense block structure in our DRDCN network for surrogate modeling of solute transport in 
media with non-Gaussian conductivities. The network architecture is shown in Figure~\ref{fig:net4surrogate}c. 
For 2-D or 3-D images (fields), the 2-D and 3-D operations (i.e., BN and Conv) implemented in the PyTorch software are respectively used in the network. The network contains 
four residual-in-residual dense blocks and the feature maps are to go through a coarsen-to-refine process. A convolutional layer 
is first employed to extract feature maps from the raw input image. The obtained features are then passed through the 
residual-in-residual dense blocks and the transition convolutional layers for downsampling/upsampling of the feature maps.
We arrange the position of the four residual-in-residual dense 
blocks in the network with a layout of $(1,2,1)$ (Figure~\ref{fig:net4surrogate}c) to encourage the information flow through the coarse feature maps. 
That is, one block is placed in the coarsening part; two adjacent blocks are placed in the most central part; and another one block is placed in the refining part.  
An additional level of residual learning is implemented to the stacked residual-in-residual dense blocks, resulting in a three-level residual learning structure in the network. In DRDCN, the Mish activation is used unless otherwise stated.

\subsection{CAAE for Parameterization of Non-Gaussian Random Fields}\label{sec:CAAE}
We propose to parameterize non-Gaussian conductivity fields with multimodal distributions using a CAAE network. Without loss of generality, here we use $\mathbf{x}$ to denote the log-conductivity field. Adversarial 
autoencoder is a probabilistic autoencoder that uses the GAN framework as a variational inference algorithm~\cite{Makhzani2016}. 
The original adversarial autoencoder framework is composed of fully-connected layers~\cite{Makhzani2016}, making it increasingly 
difficult to train as the network gets deeper due to a large number of trainable parameters. To resolve this issue, we develop a 
CAAE framework based on convolutional layers to leverage their sparse-connectivity and parameter-sharing properties as well 
as robust capability in image-like data processing~\cite{Goodfellow-et-al-2016,laloy2018,mo2019UQ,shen2018}.

\subsubsection{Generative Adversarial Networks}\label{sec:GAN}
GAN~\cite{Goodfellow2014} is a framework that establishes an adversarial game between two networks: a generative network $\mathcal{G}(\cdot)$ (i.e., generator) that learns the distribution $p_{\mathrm{data}}(\mathbf x)$ over the data, and a discriminative network $\mathcal{D}(\cdot)$ (i.e., discriminator)  that computes the probability that a sample $\mathbf x$ is sampled from 
$p_{\mathrm{data}}(\mathbf x)$, rather than generated by the generator. The generator maps the low-dimensional latent vector $\mathbf z$ from 
the prior distribution $p(\mathbf{z})$ to the data space. The discriminator is trained to maximize the probability of 
distinguishing the real samples from the generated (fake) samples. The generator is simultaneously trained to maximally 
fool the discriminator into assigning a higher probability to the generated samples by leveraging the feedback from 
the discriminator. Mathematically, the adversarial game translates into the following minimization-maximization 
loss function~\cite{Goodfellow2014}
\begin{linenomath*}
\begin{equation}\label{eq:min-max-loss}
    \min _{\mathcal{G}} \max _{\mathcal{D}} \mathbb{E}_{\mathbf{x} \sim p_{\mathrm{data}}(\mathbf{x})}\big[\log \mathcal{D}(\mathbf{x})\big]+\mathbb{E}_{\mathbf{z} \sim p(\mathbf{z})}\Big\{\log\big[1-\mathcal{D}[\mathcal{G}(\mathbf{z})]\big]\Big\}.
\end{equation}
\end{linenomath*}
In practice, the generator and discriminator are usually trained in alternating steps: (1) train the discriminator to improve 
its discriminative capability; (2) train the generator to improve the quality of the generated samples so as to fool the discriminator.

\subsubsection{Adversarial Autoencoder}\label{sec:AAE}
An autoencoder is a framework that learns a low-dimensional representation $\mathbf z$ 
(referred to as latent codes or latent variables) of a sample $\mathbf x$ in the data and generates from the codes a 
reconstruction $\hat{\mathbf x}$ that closely matches $\mathbf x$. It consists of two networks: an encoder to learn a 
mapping from $\mathbf x$ to $\mathbf z$ and a decoder to learn a mapping from $\mathbf z$ to $\hat{\mathbf x}$. To create 
a generative framework, in adversarial autoencoder~\cite{Makhzani2016} a constraint is added on the encoder that forces it to generate latent codes $\mathbf z$ 
that roughly follow a desired distribution, like a standard normal distribution $\mathcal{N}(\mathbf{0},\mathbf{I})$ 
used in the present study. The decoder is then trained to generate samples with features being consistent with those 
found in the training data given any sample $\mathbf z\sim\mathcal{N}(\mathbf{0},\mathbf{I})$ as input, resulting in a generative model. 

Mathematically, let $q(\mathbf{z}|\mathbf x)$ be the encoding distribution, $q(\mathbf{x}|\mathbf{z})$ be  the 
decoding distribution, and $p(\mathbf{z})$ be the distribution that we want the latent variables $\mathbf{z}$ to follow. The adversarial autoencoder looks for a generative model  
\begin{linenomath*}
\begin{equation}\label{eq:generative-model}
    p(\mathbf{x})=\int p(\mathbf{x}|\mathbf{z})p(\mathbf{z}) d \mathbf{z}.
\end{equation}
\end{linenomath*}
The adversarial autoencoder~\cite{Makhzani2016} is very similar to the VAE~\cite{Kingma2014} in the sense that 
in both a latent representation is obtained with a desired distribution. Thus we follow 
the formulation in VAE to finally introduce the adversarial autoencoder. In VAE, the generative model is obtained 
via minimizing the upper-bound of the negative log-likelihood
\begin{linenomath*}
\begin{equation}\label{eq:VAE-loss}
        \mathbb{E}_{\mathbf{x} \sim p_{\mathrm{data}}(\mathbf{x})}\big[-\log p(\mathbf{x})\big]< \mathbb{E}_{\mathbf{x} \sim p_{\mathrm{data}}(\mathbf{x})}\Big\{
        \mathbb{E}_{\mathbf{z}\sim q(\mathbf{z} | \mathbf{x})}\big[-\log p(\mathbf{x} | \mathbf{z})\big]+\mathrm{KL}\big[q(\mathbf{z} | \mathbf{x}) \| p(\mathbf{z})\big]
    \Big\}.
\end{equation}
\end{linenomath*}
The first term on the right side quantifies the reconstruction quality and the second term is the Kullback-Leibler 
divergence measuring the difference between two distributions. 

\begin{figure}[h!]
    \centering
    \includegraphics[width=\textwidth]{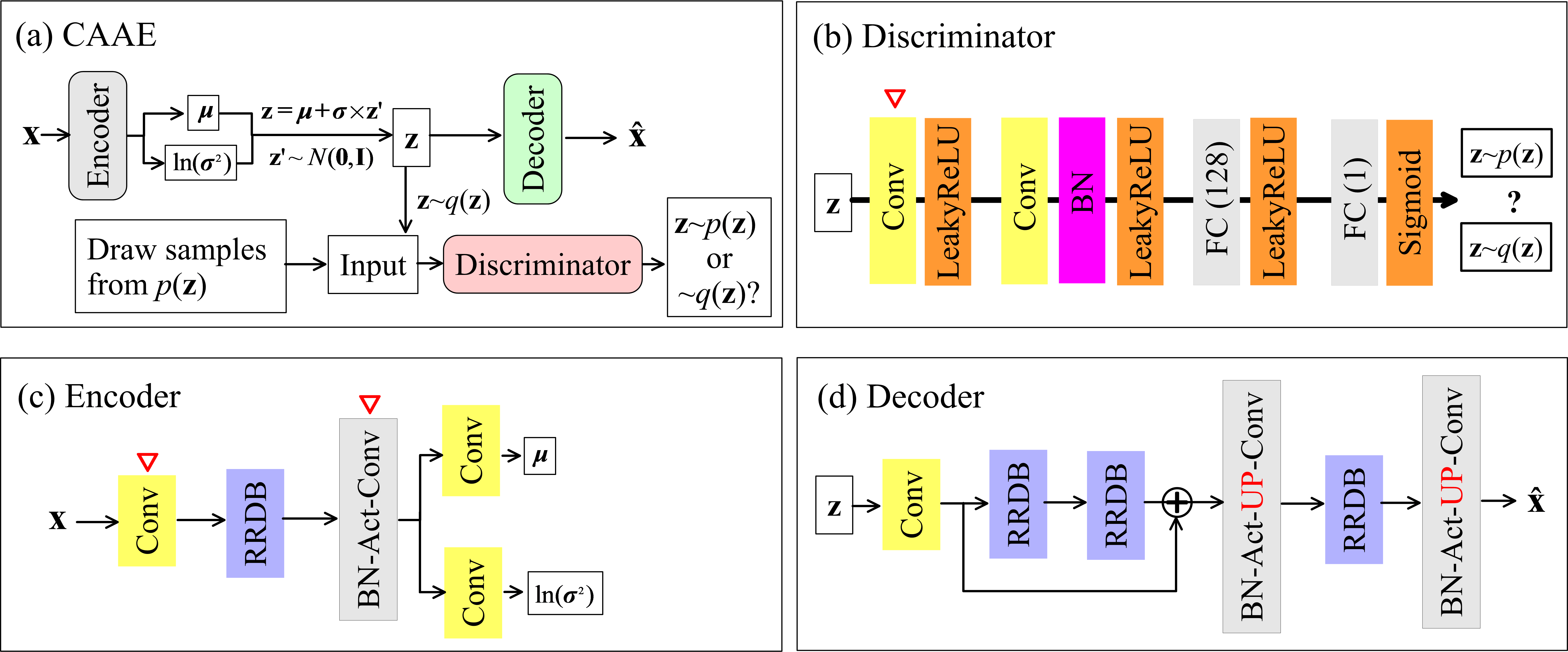}
    \caption{(a) Illustration of a convolutional adversarial autoencoder (CAAE), which is composed of a discriminator (b), an encoder (generator) (c), 
    and a decoder (d). The discriminator is a stack of two convolutional layers followed by two fully-connected (FC) layers with $128$ and $1$ neurons, 
    respectively. The feature map size is halved in Conv ($\bigtriangledown$) with stride $2$ 
    in the discriminator and encoder, and doubled in the decoder by using the nearest upsampling (UP) operation.}
    \label{fig:CAAE} 
\end{figure}

In adversarial autoencoders, an adversarial training procedure instead of the Kullback-Leibler divergence is  used to
encourage an aggregated posterior distribution $q(\mathbf{z})$, instead of $q(\mathbf{z}|\mathbf{x})$, to 
match $p(\mathbf{z})$, where $q(\mathbf{z})$ is defined as~\cite{Makhzani2016}
\begin{linenomath*}
\begin{equation}
    q(\mathbf{z})=\int_{\mathbf{x}} q(\mathbf{z} | \mathbf{x}) p_{\mathrm{data}}(\mathbf{x}) d \mathbf{x}.
\end{equation}
\end{linenomath*}
An illustration of the adversarial autoencoder is depicted in Figure~\ref{fig:CAAE}a. In the encoding path, the input $\mathbf x$ is 
fed into the encoder which outputs two low-dimensional vectors of means $\boldsymbol{\mu}$ and log-variances $\ln(\boldsymbol{\sigma}^2)$ of the latent variables $\mathbf{z}$. 
Then a vector $\mathbf{z}'$ is randomly drawn from $\mathcal{N}(\mathbf{0},\mathbf{I})$ and rescaled to produce the codes $\mathbf{z}=\boldsymbol{\mu}+\boldsymbol{\sigma}\times\mathbf{z}'$, where $\times$ denotes element-wise 
multiplication.  The decoder takes $\mathbf{z}$ as input to eventually generate $\hat{\mathbf{x}}$. Meanwhile, the discriminator of the 
adversarial network accepts input from the latent codes generated by the encoder or the prespecified 
distribution $p(\mathbf{z})$ to discriminatively predict whether the input arises from the encoding codes (fake sample) or $p(\mathbf{z})$ 
(real sample). Note that the adversarial network here differs slightly from the vanilla GAN framework~\cite{Goodfellow2014}, 
in which the generator generates the sample $\mathbf{x}$, and the discriminator discriminates $\mathbf{x}$ whether it is from 
the generator or the data. 

The encoder (which is also the generator $\mathcal{G}(\cdot)$ of the adversarial network), decoder, and discriminator $\mathcal{D}(\cdot)$ of the 
adversarial autoencoder are trained jointly in two phases for each iteration: the reconstruction phase and the regularization phase~\cite{Makhzani2016}. 
In the reconstruction phase, the encoder (generator) and decoder are updated using the following loss function:
\begin{linenomath*}
\begin{equation}\label{eq:G-loss}
    \mathcal{L}_{\mathrm{ED}}=\mathcal{L}_{\mathrm{Rec}}+w\mathcal{L}_\mathcal{G},
\end{equation}
\end{linenomath*}
where $\mathcal{L}_{\mathrm{Rec}}$ is the reconstruction error which in this study is taken as the $L_1$ loss:
\begin{linenomath*}
\begin{equation}
    \mathcal{L}_{\mathrm{Rec}}=\frac{1}{N}\sum_{i=1}^{N}||\mathbf{x}_i-\hat{\mathbf{x}_i}||_1,
\end{equation}
\end{linenomath*}
and $\mathcal{L}_\mathcal{G}$ measures the generator's ability to fool the discriminator and has the form
\begin{linenomath*}
\begin{equation}
    \mathcal{L}_\mathcal{G}=-\frac{1}{N}\sum_{i=1}^{N}\log\big\{\mathcal{D}[\mathcal{G}(\mathbf{x}_i)]\big\}.
\end{equation}
\end{linenomath*}
Here, $w$ is a weight factor balancing the two losses and a value of $w=0.01$ is used, $\hat{\mathbf{x}}_i$ is the reconstruction of sample $\mathbf{x}_i$, and $N$ is the number of training samples. 
In the regularization phase, the discriminator is updated based on the loss function 
\begin{linenomath*}
\begin{equation}\label{eq:D-loss}
    \mathcal{L}_\mathcal{D}=-\frac{1}{N}\sum_{i=1}^{N}\Big\{\log\big[\mathcal{D}(\mathbf{z}_i)\big]+\log{\big[1-\mathcal{D}[\mathcal{G}(\mathbf{x}_i)\big]}\Big\},
\end{equation}
\end{linenomath*}
to distinguish the real sample from $p(\mathbf{z})$ (i.e. $\mathbf{z}_i$) from the fake sample $\mathcal{G}(\mathbf{x}_i)$ produced by the generator. 

Such adversarial training process with the loss functions defined in equations~(\ref{eq:G-loss}) and~(\ref{eq:D-loss}) forces $\hat{\mathbf{x}}$ to 
closely match $\mathbf{x}$ and $q(\mathbf{z})$ to gradually approach $p(\mathbf{z})$ (i.e., $\boldsymbol{\mu}\rightarrow{\mathbf{0}}$ 
and $\boldsymbol{\sigma}^2\mathbf{I}\rightarrow{\mathbf{I}}$), respectively. After training, the decoder will define a generative 
model $p(\mathbf{x})$ that given an arbitrary input $\mathbf{z}\sim\mathcal{N}(\mathbf{0},\mathbf{I})$ can 
generate a new realization of sample $\hat{\mathbf{x}}$ with features similar to those in the data used for training (Figure~\ref{fig:CAAE}d).

\subsubsection{CAAE Networks Based on Residual-in-Residual Dense Blocks}\label{sec:CAAE-arch}
We also adopt the residual-in-residual dense block structure shown in Figure~\ref{fig:net4surrogate}b in the encoder and decoder 
of the CAAE network. The encoder (Figure~\ref{fig:CAAE}c) is similar to the coarsening part of the DRDCN network (Figure~\ref{fig:net4surrogate}c). The encoder has two additional convolutional layers to respectively output the means $\boldsymbol{\mu}$ and log-variances $\ln{(\boldsymbol{\sigma}^2)}$. The decoder (Figure~\ref{fig:CAAE}d) is similar to the refining part of the DRDCN network.  The decoder has an additional convolutional layer to 
extract feature maps from the codes $\mathbf{z}$. The ReLU activation is used in the encoder and decoder. Inspired by~\citeA{Ledig2017CVPR}, the discriminator is a stack of two convolutional layers followed by two fully-connected layers with $128$ and $1$ neurons, 
respectively. The leaky ReLU activation with a slope of $0.2$ is used in the discriminator and the 
sigmoid activation is used in the last layer to output a probability value between $0$ and $1$.

\subsection{The CAAE-DRDCN-ILUES inversion framework}\label{sec:CAAE-DRDCN-ILUES}
We incorporate the CAAE parameterization method and the DRDCN surrogate method into ILUES to formulate an 
efficient inversion scheme for estimation of a non-Gaussian conductivity field of solute transport models. 
The integrated methodology is denoted as CAAE-DRDCN-ILUES hereinafter and is summarized in Algorithm~\ref{algor:CAAE-DRDCN-ILUES}. 
Notice that in this method, the surrogate model is used in Algorithms~\ref{algor:ilues-update} and~\ref{algor:CAAE-DRDCN-ILUES} 
to substitute the forward model. After parameterization, the uncertain parameters to be estimated are the latent variables $\mathbf{z}$. 
The log-conductivity field is estimated with the following procedure: (1)  start with an initial latent code ensemble drawn 
from $\mathcal{N}(\mathbf{0},\mathbf{I})$, (2) the corresponding log-conductivity fields are generated next using the CAAE's decoder, (3) the surrogate model is evaluated to obtain the predicted initial output ensemble, (4) the latent code and output ensembles are repeatedly updated for $N_{\rm{iter}}$ iterations using 
Algorithm~\ref{algor:ilues-update} based on the current latent code and output ensembles. In Algorithm~\ref{algor:ilues-update}, the input to the surrogate model to produce output predictions is the log-conductivity field, which is generated by the CAAE's decoder given the latent codes as input. The posterior log-conductivity fields are obtained from the decoder using the last latent code ensemble as input.

\begin{algorithm}[htb]
\caption{The CAAE-DRDCN-ILUES inversion framework for estimation of the log-conductivity field $\mathbf{x}$. The log-conductivity field realizations are generated using the CAAE's \texttt{decoder} given the latent variables $\mathbf{z}$ as input. CAAE:  convolutional adversarial autoencoder. \texttt{DRDCN}: deep residual dense convolutional network. ILUES: iterative local updating ensemble smoother.}
\label{algor:CAAE-DRDCN-ILUES}
\begin{algorithmic}[1]

\Require Measurements $\bm{d}$, iteration number $N_{\rm iter}$, ensemble size $N_e$, trained \texttt{decoder}, trained \texttt{DRDCN}.

\State Generate the initial input ensemble $\mathbf{Z}^0=[\mathbf{z}_1^0,\ldots,\mathbf{z}_{N_e}^0]$ from $\mathcal{N}(\mathbf{0},\mathbf{I})$. 
\State Generate the initial log-conductivity field ensemble $\hat{\mathbf{X}}^0=[\hat{\mathbf{x}}_1^0,\ldots,\hat{\mathbf{x}}_{N_e}^0]$ via $\hat{\mathbf{x}}_i^0=\texttt{decoder}(\mathbf{z}_i^0)$.
\State Obtain the initial output ensemble $\hat{\mathbf{D}}^0=\big[\hat{f}(\hat{\mathbf{x}}_1^0),\ldots,\hat{f}(\hat{\mathbf{x}}_{N_e}^0)\big]$ via $\hat{f}(\hat{\mathbf{x}}_i^0)=\texttt{DRDCN}(\hat{\mathbf{x}}_i^0)$.

\For{$n=0,\ldots,(N_{\rm iter}-1)$} \Comment Iterative data assimilation
   
  \State Obtain $\mathbf{Z}^{n+1}=[\mathbf{z}_1^{n+1},\ldots,\mathbf{z}_{N_e}^{n+1}]$ and  $\hat{\mathbf{D}}^{n+1}=\big[\hat{f}(\hat{\mathbf{x}}_1^{n+1}),\ldots,\hat{f}(\hat{\mathbf{x}}_{N_e}^{n+1})\big]$ based on $\{\mathbf{Z}^n,\hat{\mathbf{D}}^n,\bm{d}\}$ using Algorithm~\ref{algor:ilues-update}.

\EndFor
\State \textbf{end for}
\State $\hat{\mathbf{X}}^{N_{\rm{iter}}}=[\hat{\mathbf{x}}_1^{N_{\rm{iter}}},\ldots,\hat{\mathbf{x}}_{N_e}^{N_{\rm{iter}}}]$, where $\hat{\mathbf{x}}_i^{N_{\rm{iter}}}=\texttt{decoder}(\mathbf{z}_i^{N_{\rm{iter}}})$. 
\State \textbf{return} $\hat{\mathbf{X}}^{N_{\rm iter}}$, $\hat{\mathbf{D}}^{N_{\rm{iter}}}$  \Comment{The final log-conductivity field and output ensembles}
\end{algorithmic}
\end{algorithm}

\section{Application}\label{sec:application}

\subsection{Solute Transport Models}\label{sec:conceptual-model}

The performance of the proposed method is illustrated using 2-D and 3-D solute transport modeling with random conductivity fields that have non-Gaussian heterogeneity patterns.

\subsubsection{2-D Model}
\begin{figure}[h!]
    \centering
    \includegraphics[width=\textwidth]{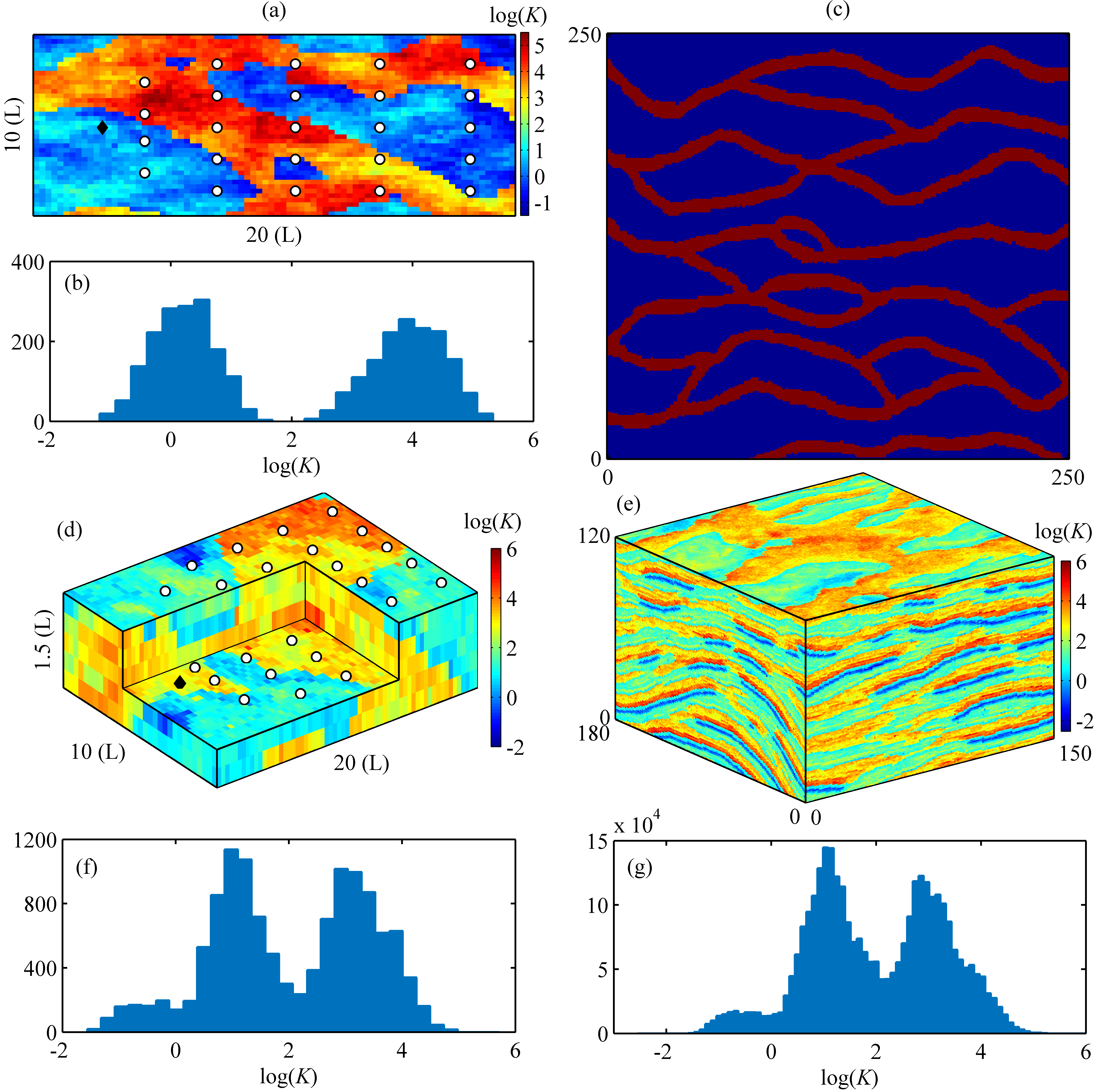}
    \caption{Reference log-conductivity conductivity fields of the 2-D (a) and 3-D (d) solute transport models considered in the inverse problems. The diamond and $24$ circles represent the projections of the solute source and observation locations, respectively, in the horizontal plane. Histograms of the log-conductivities in (a) and (d) are shown in (b) and (f), respectively. (c) The $250\times250$ training image used to generate 2-D binary facies fields. (e) The $120\times 180\times150$ training image used to generate 3-D log-conductivity fields. (g) Histogram of the log-conductivities in (e).}
    \label{fig:refK} 
\end{figure}

 The first test case considers 2-D solute transport within a channelized aquifer.  As shown in Figure~\ref{fig:refK}a, the horizontal domain has a size of $10~\rm{(L)}\times 20~\rm{(L)}$ and is uniformly discretized into $H\times W=40\times80=3,200$ gridblocks. The left and right 
boundaries are assumed to be constant head boundaries with heads of $1~\rm{(L)}$ and $0~\rm{(L)}$, respectively. 
No-flow boundary condition is imposed on the upper and lower boundaries. An instantaneous source with a concentration of $100$ (ML$^{-3}$) is released from the location $x=3$~(L) and $y=5$~(L) at the initial time. The porosity and dispersivities 
are assumed to be known with constant values of $\phi=0.25$, $\alpha_L=1.0$~(L) and $\alpha_T=0.1$~(L), respectively.

The non-Gaussian conductivity field in this case has a channelized and bimodal pattern with heterogeneous conductivity within each facies. The conductivity realizations are generated by the following procedure: First, 
a binary facies field is generated using the {SNESIM} code~\cite{Strebelle2002} with a training image shown in Figure~\ref{fig:refK}c; then 
we populate each facies with log-conductivity values from two independently generated Gaussian random fields with a $L_2$ norm exponential covariance function:
\begin{linenomath*}
\begin{equation}
    C(\bm{s},\bm{s}')=\sigma_{\log{(K)}}^2\exp{\left(-\sqrt{\left(\frac{s_1-s'_1}{\lambda_1}\right)^2+\left(\frac{s_2-s'_2}{\lambda_2}\right)^2}\right)},
    \label{eq:cov}
\end{equation}
\end{linenomath*}
where $\bm{s}=(s_1,s_2)$ and $\bm{s}'=(s'_1,s'_2)$ denote two arbitrary spatial locations, $\sigma_{\log{(K)}}^2$ is the variance, and $\lambda_1$ 
and $\lambda_2$ are the correlation lengths along the $x-$ and $y-$axes, respectively. The means of Gaussian random fields corresponding to 
the high-conductivity channels and the low-conductivity non-channel medium are $4$ and $0$, respectively, while their variances and correlation 
lengths are taken the same with $\sigma_{\log{(K)}}^2=0.5$, $\lambda_1=4$~(L), and $\lambda_2=2$~(L).

\subsubsection{3-D Model}
The second test case considers  solute transport in a 3-D confined aquifer with a size of $1.5~(\rm{L})\times 10~(\rm{L})\times 20~(\rm{L})$ as depicted in Figure~\ref{fig:refK}d. 
The domain is uniformly discretized into $D\times H\times W=6\times 32\times 64=12,288$ gridblocks. Similar to the 2-D case, the left and right boundaries are assumed to be 
constant head boundaries with heads of $1~\rm{(L)}$ and $0~\rm{(L)}$, respectively. No-flow boundary condition is imposed on the boundaries in the direction 
perpendicular to the $y$-axis. An instantaneous source with a concentration of $100$ (ML$^{-3}$) is released from 
the location $x=3$~(L) and $y=5$~(L) at the initial time. The porosity and dispersivities are assumed to be 
known with constant values of $\phi=0.25$, $\alpha_L=1.0$~(L), $\alpha_T=0.1$~(L), and $\alpha_V=0.01$~(L), respectively. 
The conductivity fields for this 3-D model are obtained by randomly cropping $6\times 32\times 64$ patches from a $120\times 180\times 150$ training image depicted in Figure~\ref{fig:refK}e (available at \url{http://www.trainingimages.org/training-images-library.html}). 
The conductivity heterogeneity pattern is different from that in the 2-D channelized field in the sense that the distribution of log-conductivities in this training image is trimodal with two major peaks around $1.0$ and $3.0$ and one minor peak around $-0.6$ (Figure~\ref{fig:refK}g).

\subsection{Synthetic Observations}\label{sec:synthetic-obs}
As it will be shown in section~\ref{sec:CAAE-parameterized}, the CAAE-generated conductivity fields have higher regularity/smoothness 
than the original fields. In the inverse problem, the synthetic observations were obtained by running the forward model with 
an original conductivity field rather than a smoothed conductivity field generated by CAAE. The generation of synthetic observations 
here aims to mimic a real scenario, where data would be obtained by field measurements. The randomly generated 2-D and 3-D reference 
log-conductivity fields are depicted in Figures~\ref{fig:refK}a and~\ref{fig:refK}d, respectively. Note that they are distinct 
from the training log-conductivity samples of CAAE and DRDCN networks. The two reference fields are both non-Gaussian with a bimodal distribution (the 2-D case, Figure~\ref{fig:refK}b) and a trimodal distribution (the 3-D case, Figure~\ref{fig:refK}f), respectively. In the 2-D case, the 
concentration at $t=[3,5,7,9,11]$~(T) and the hydraulic head are collected at $24$ measurement locations (Figure~\ref{fig:refK}a), resulting in $144$ observations. 
In the 3-D case, the concentration at $t=[4,6,8]$~(T) and the hydraulic head are collected at six depths of $24$ measurement 
locations (Figure~\ref{fig:refK}d), resulting in $576$ observations. The synthetic observations were 
corrupted with $5\%$ independent Gaussian random noise to the data generated by the reference model. Additionally, we do not use 
any conditioning data (i.e., measurements) of the conductivity, resulting in a rather challenging inverse problem.

\subsection{Networks Design and Training}

\subsubsection{CAAE Network Design and Training}\label{sec:CAAE-design-train}
The architecture of the CAAE network is shown in Figure~\ref{fig:CAAE} and detailed in section~\ref{sec:CAAE-arch}. The encoder includes two downsampling layers which halves the feature map size via convolution with a stride of $2$. Correspondingly, the two upsampling layers in the decoder double the feature 
map size to recover the output image size. The network consists of $71$ layers, including $69$ convolutional layers 
(mostly arising from the four residual-in-residual dense blocks that each contains $15$ convolutional layers) 
and $2$ fully-connected layers (in the discriminator). The kernel size in all convolutional layers is $3$, and the stride in the convolutional 
layers that keep the same feature map size and halve the size is $1$ and $2$, respectively.

In the 2-D case, a training set with $40,000$ realizations of the log-conductivity field is generated.
We also generate another $4,000$ test realizations to evaluate the network's performance. In the 3-D case, we generate the log-conductivity realizations 
by cropping the $120\times 180\times 150$ training image shown in Figure~\ref{fig:refK}e. The original training image is split into two images, with the lower $105\times 180\times 150$ part and the upper $15\times 180\times 150$ part being used to generate the training and test datasets, respectively. The two images are both flipped 
along the three dimensions via the \texttt{flip} operation implemented in MATLAB to augment the data, each resulting in four training images. The log-conductivity realizations with size $6\times 32\times 64$ are then generated via cropping the training images using a stride of $(2,6,10)$. In this way, we obtain $37,600$ training samples and $3,000$ test samples. The data augmentation strategy, which artificially creates new training data from existing training data via specific operations (e.g., flip, shift, and rotation), is commonly utilized to obtain improved performance~\cite{Krizhevsky2012_4824}. Note that these operations may lead to correlation between the original and resulting samples, thus the training and test data are generated separately as mentioned above to avoid potential over-optimistic results when assessing the performance.
The reference log-conductivity fields used in section~\ref{sec:synthetic-obs} to generate the synthetic observations are randomly selected from the test sets.
The loss functions for network training are defined in equations~(\ref{eq:G-loss}) and~(\ref{eq:D-loss}). 
The network is trained on a NVIDIA GeForce GTX $1080$ Ti X GPU for $50$ epochs using the Adam optimizer~\cite{kingma2014-adam} with a learning rate of $2\times 10^{-4}$ and a batch size of $64$. The training required about 1.7~h and 13.1~h in the 2-D and 3-D cases, respectively.

\subsubsection{DRDCN Network Design and Training}
The architecture of the DRDCN network is shown in Figure~\ref{fig:net4surrogate} and detailed 
in section~\ref{sec:DRDCN-arc}. The network is fully-convolutional and contains $64$ convolutional layers without any fully-connected layers. 
The kernel size in all convolutional layers is $3$, and the stride in the convolutional 
layers that keep the same feature map size and halve the size is $1$ and $2$, respectively. The softplus activation is used in the output layer for the concentration to ensure nonnegative predictions. Since the hydraulic head varies between $0$ and $1$ in both cases, the sigmoid activation is used in the output layer for the hydraulic head. 

The concentration at different time steps (i.e., $t=[3,5,7,9,11]$~(T) and $t=[4,6,8]$~(T) in the 2-D and 3-D cases, respectively) and the hydraulic head are collected as the observations in the inverse problem. 
Thus, the concentration fields at these time steps and the hydraulic head field are treated as the output channels of the network. 
There is one single input channel to the network which is the original log-conductivity field generated by following the 
procedure presented in section~\ref{sec:conceptual-model}. In both cases, we generate four training sets with $N=1,000$, $2,000, 3,000$, and $4,000$ 
samples to evaluate the convergence of the network approximation errors with respect to   the training sample size. 
The approximation accuracy is assessed using $N_{\rm{test}}=1,000$ randomly generated test samples. Note that since the 3-D log-conductivity realizations are obtained via cropping the training image which may lead to data correlation issue (see section~\ref{sec:CAAE-design-train}), the log-conductivity samples in the training and test data are randomly selected from the training and test datasets of the CAAE network, respectively. 
The accuracy is measured using the coefficient 
of determination ($R^2$), the root-mean-square error (RMSE), and the structural similarity index (SSIM) metrics. The $R^2$ and RMSE metrics are defined as
 \begin{linenomath*}
 \begin{equation}
     R^2=1-\frac{\sum_{i=1}^{N_{\rm{test}}}||\mathbf y_i-\hat{\mathbf y}_i||_2^2}{\sum_{i=1}^{N_{\rm{test}}}||\mathbf y_i-\bar{\mathbf y}||_2^2},
 \end{equation}
 \end{linenomath*}
 and
 \begin{linenomath*}
 \begin{equation}
     \text{RMSE}=\sqrt{\frac{1}{N_{\rm{test}}}\sum_{i=1}^{N_{\rm{test}}}||\mathbf{y}_i-\hat{\mathbf{y}}_i||_2^2},
 \end{equation}
 \end{linenomath*}
 respectively, where $\mathbf{y}$ denotes the simulated outputs, $\hat{\mathbf{y}}$ is the network predictions, and $\bar{\mathbf{y}}=1/N_{\text{test}}\sum_{i=1}^{N_{\text{test}}}\mathbf{y}_i$. The SSIM metric measures the structural similarity between two images (fields) and is calculated over local windows of the image~\cite{Wangetal2004}
\begin{linenomath*}
\begin{equation}
    \mathrm{SSIM}(\mathbf{u}, \mathbf{v})=\frac{\left(2 \mu_{\mathbf{u}} \mu_{\mathbf{v}}+c_{1}\right)\left(2 \sigma_{\mathbf{u}\mathbf{v}}+c_{2}\right)}{\left(\mu_{\mathbf{u}}^{2}+\mu_{\mathbf{v}}^{2}+c_{1}\right)\left(\sigma_{\mathbf{u}}^{2}+\sigma_{\mathbf{v}}^{2}+c_{2}\right)},
\end{equation}
\end{linenomath*}
where $\mathbf{u}$ and $\mathbf{v}$ are two windows with size $11\times11$ in the real and predicted images, respectively, $\mu_{\mathbf{u}}$ ($\mu_{\mathbf{v}}$) and $\sigma_{\mathbf{u}}^{2}$ ($\sigma_{\mathbf{v}}^{2}$) are the mean and variance values of window $\mathbf{u}$ ($\mathbf{v}$), $\sigma_{\mathbf{uv}}$ denotes the covariance between $\mathbf{u}$ and $\mathbf{v}$, and $c_1=0.01$ and $c_2=0.03$ are two constants~\cite{Wangetal2004}. Note that the SSIM metric is designed for 2-D images. A 3-D image with size $D\times H\times W$  is treated as $D$ images with size $H\times W$ when computing the SSIM metric. A $R^2$ score value and a SSIM value approaching $1.0$ and a lower RMSE value suggest better surrogate quality.

The network is trained using a regularized $L_1$ loss function:
\begin{linenomath*}
\begin{equation}\label{eq:l1-loss}
    \mathcal{L}=\frac{1}{N}\sum_{i=1}^{N}||\mathbf{y}_i-\hat{\mathbf{y}_i}||_1+\frac{w_d}{2}\boldsymbol{\theta}^{\top}\boldsymbol{\theta},
\end{equation}
\end{linenomath*}
where $\boldsymbol{\theta}$ denotes all the network trainable parameters and $w_d=1\times10^{-5}$ is a regularization coefficient. 
The network is trained on a NVIDIA GeForce GTX $1080$ Ti X GPU for $200$ epochs in the 2-D case and $300$ epochs in the 3-D case using the Adam optimizer~\cite{kingma2014-adam} with an initial learning rate of 
$5\times 10^{-3}$ and a batch size of $32$. We also use a learning rate scheduler which drops ten times on plateau during training. The training required about 0.3-1.2~h and 2.5-10.5~h in the 2-D and 3-D cases, 
respectively, as the training set size varies from $1,000$ to $4,000$.

\section{Results and Discussion}\label{sec:results&discussion}
In this section, we first illustrate the performance of the CAAE and DRDCN networks in parameterization of non-Gaussian random fields 
and in surrogate modeling of the solute transport models, respectively. After that, the inversion results obtained 
from the CAAE-DRDCN-ILUES framework are compared to those obtained from the CAAE-ILUES framework without surrogate modeling.

\begin{figure}[h!]
    \centering
    \includegraphics[width=.8\textwidth]{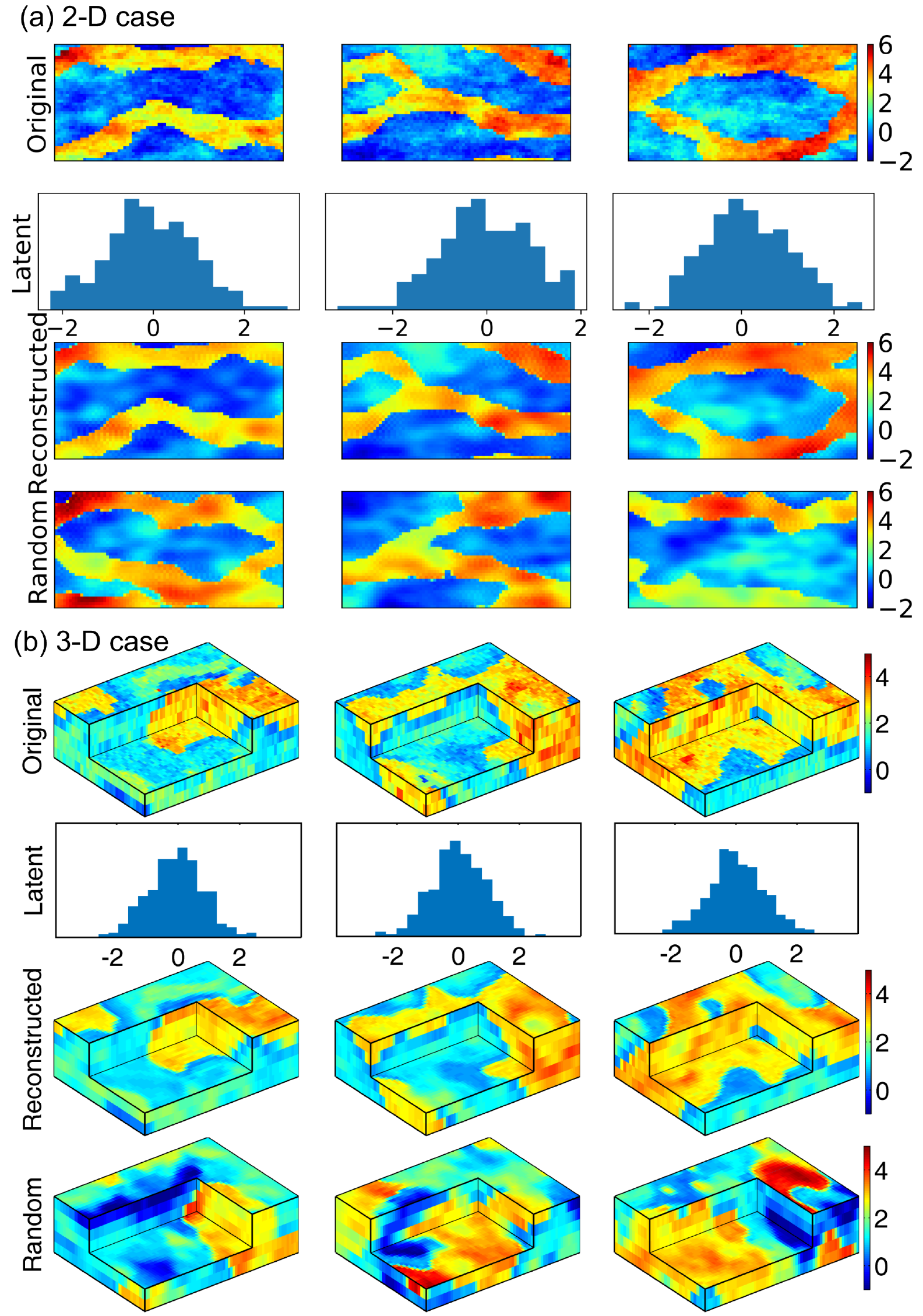}
    \caption{The CAAE network's reconstructions (row 3) for 2-D (a) and 3-D (b) log-conductivity fields (row $1$) in the test sets and the corresponding histograms of the latent codes (row $2$). The reconstruction accuracy in (a) and (b) is $R^2=(0.945,0.948,0.944)$ and $(0.854,0.837,0.852)$, respectively. The fourth 
    row in (a) and (b) shows random log-conductivity fields generated by the CAAE's decoder with inputs from $\mathcal{N}(\mathbf{0},\mathbf{I})$.}
    \label{fig:CAAE-recon-realiz} 
\end{figure}

\subsection{Parameterization of Non-Gaussian Random Fields}\label{sec:CAAE-parameterized}
In the CAAE network, the latent dimensions in the 2-D and 3-D cases are set to 200 and 512, respectively. 
The parameterization results for  non-Gaussian log-conductivity fields 
are shown in Figure~\ref{fig:CAAE-recon-realiz}. It depicts the CAAE's reconstructions of three log-conductivity realizations in the test set, the corresponding histograms of the latent codes, 
and three random log-conductivity fields generated by the decoder with inputs from $\mathcal{N}(\mathbf{0},\mathbf{I})$. 
In the reconstruction process, the original log-conductivity realization is fed to the encoder to produce latent codes, based on which the reconstructed field is generated by the decoder. 
It is observed that, in both 2-D and 3-D cases with different heterogeneity patterns, the network successfully recovers the spatial distributions of the low-conductivity and 
high-conductivity regions as well as the conductivities within these regions; although the conductivity heterogeneity is smoothed compared to the original fields. This heterogeneity smoothness is attributed to the information loss during the encoding and decoding processes.
The encoding latent codes roughly follow the prior distribution $\mathcal{N}(\mathbf{0},\mathbf{I})$ 
that we imposed during training.  The reconstruction accuracy evaluated on the test sets is $R^2=0.945$ and $0.863$ in the 2-D and 3-D cases, respectively. A larger latent dimension can be used in order to preserve more heterogeneity features in the generated fields, which however may lead to higher computational costs in inverse problems. The fourth row of Figures~\ref{fig:CAAE-recon-realiz}a and~\ref{fig:CAAE-recon-realiz}b 
shows the random log-conductivity fields generated by the decoder. The results show that the decoder is able to reproduce 
log-conductivity realizations that depict similar patterns of
heterogeneity (e.g., the channel structures and the conductivity continuity within the low/high-conductivity regions) to those found in the training data. 
Therefore, the CAAE network is employed in the inversion process as the parameterization framework for  non-Gaussian conductivity fields.

\subsection{Surrogate Quality Assessment}

\begin{figure}[h!]
    \centering
    \includegraphics[width=\textwidth]{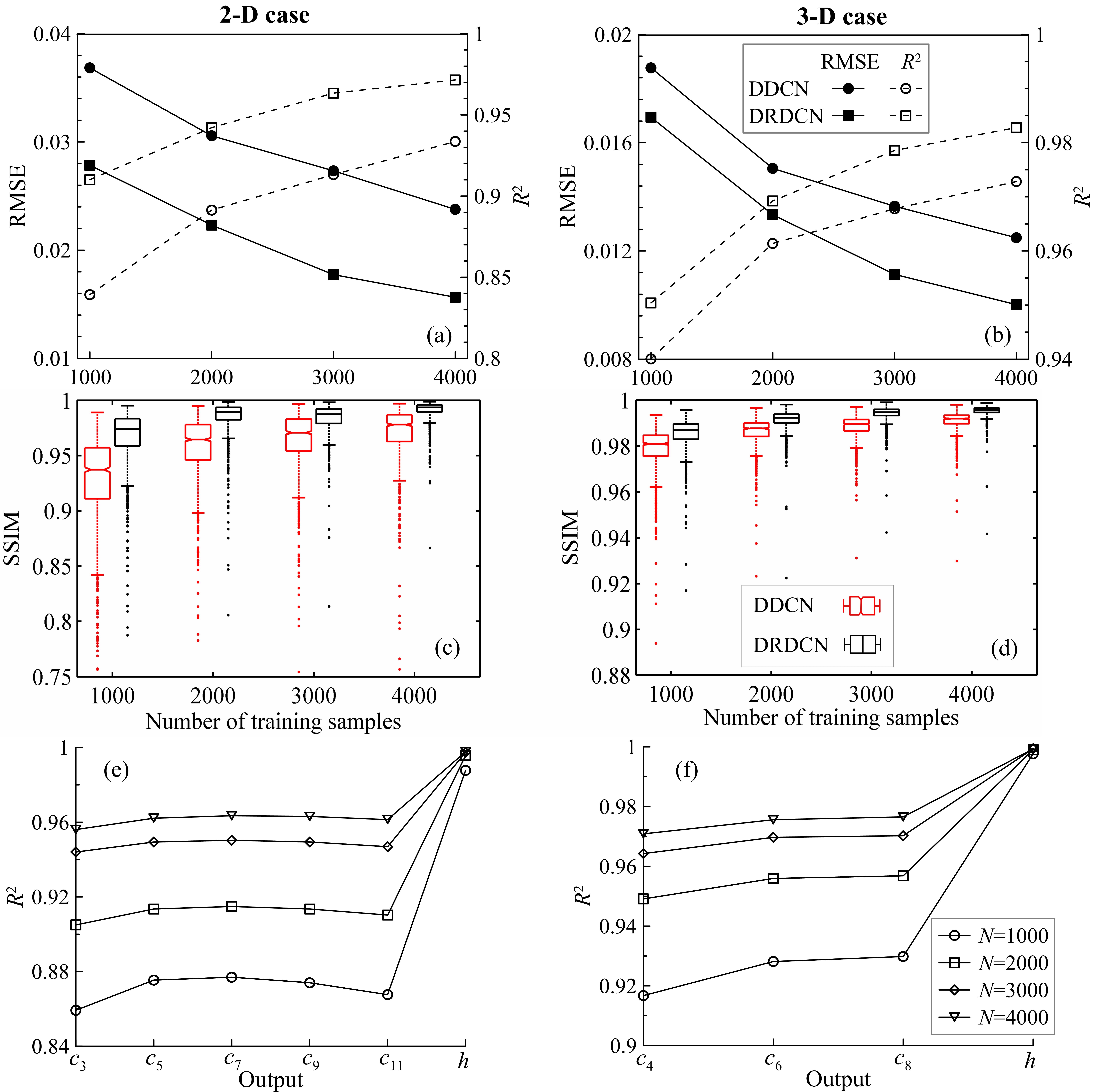}
    \caption{Comparison of the RMSEs, $R^2$ scores, and SSIMs of the DDCN and DRDCN networks evaluated on the test sets of the 2-D (left column) and 3-D cases (right column). The boxplots summarize the SSIM values evaluated on all output fields in the test sets. The metrics in (a-d) evaluate the approximation accuracy for the concentration field at all time steps considered and the hydraulic head field. (e) and (f) compare the $R^2$ scores of the DRDCN networks for the concentration ($c_t$) fields at different time $t$ (T) and the hydraulic head ($h$) field when trained using different numbers ($N$) of samples.}
    \label{fig:RMSE-R2} 
\end{figure}

To illustrate the superior performance of the proposed DRDCN network architecture against the DDCN network architecture 
employed in our previous studies~\cite{mo2019inverse,mo2019UQ,ZHU2018415,zhu2019physics} for surrogate modeling of systems 
with highly-complex input-output mappings, the DDCN network is also trained using the same training sets as those used in DRDCN. The DDCN network architecture is introduced in~\ref{appendix:DDCN}. 

\begin{figure}[h!]
    \centering
    \includegraphics[width=\textwidth]{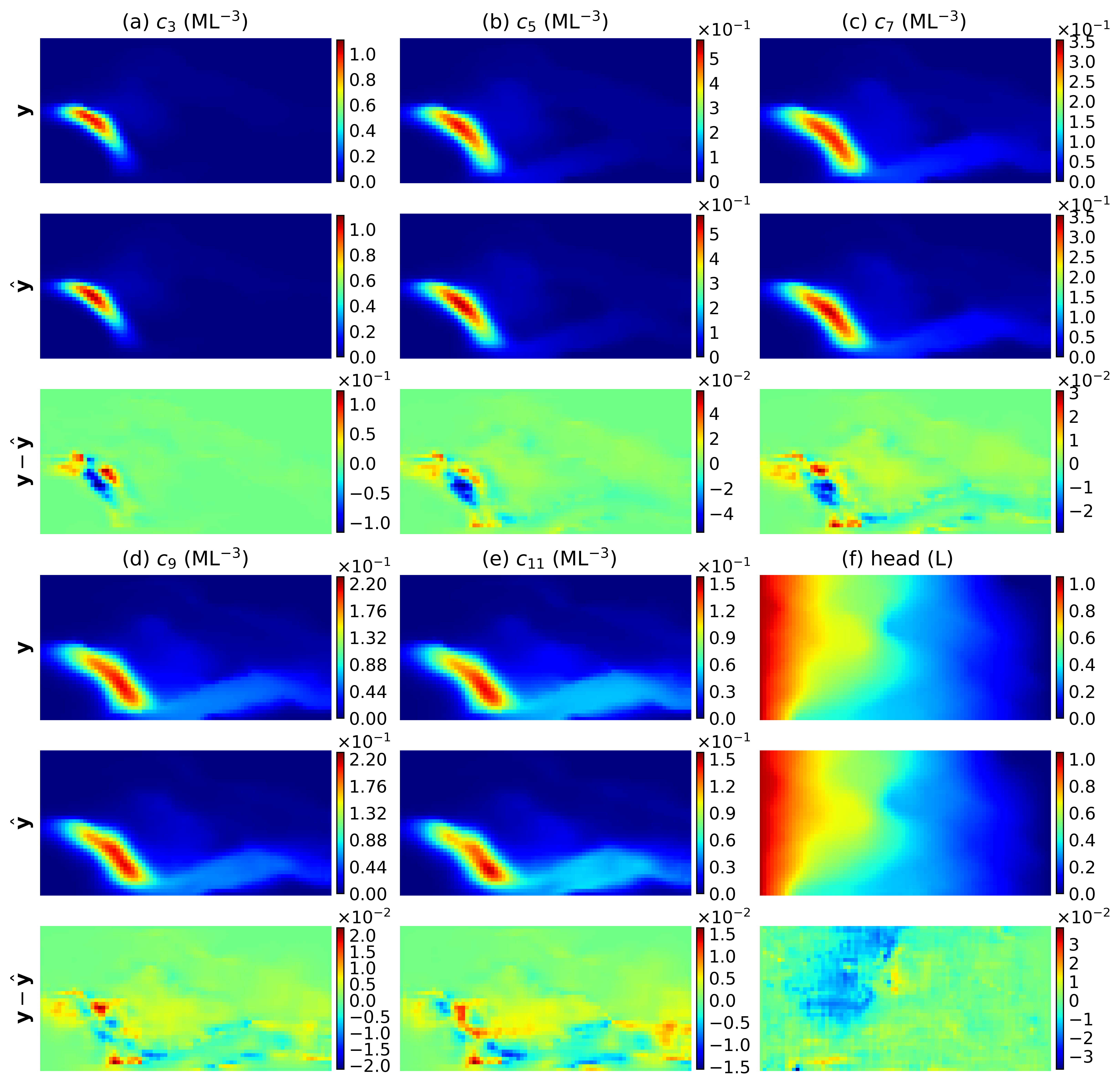}
    \caption{Comparison of the concentration $(c_t)$ fields at time $t=[3,5,7,9,11]$~(T; a-e) and the hydraulic 
    head field (f) of a random test sample predicted by the 2-D forward model ($\mathbf{y}$) and the DRDCN 
    network ($\hat{\mathbf{y}}$) trained using $4,000$ samples. ($\mathbf{y}-\hat{\mathbf{y}}$) denotes the network approximation error.}
    \label{fig:DRDCN-pred} 
\end{figure}

\begin{figure}[h!]
    \centering
    \includegraphics[width=\textwidth]{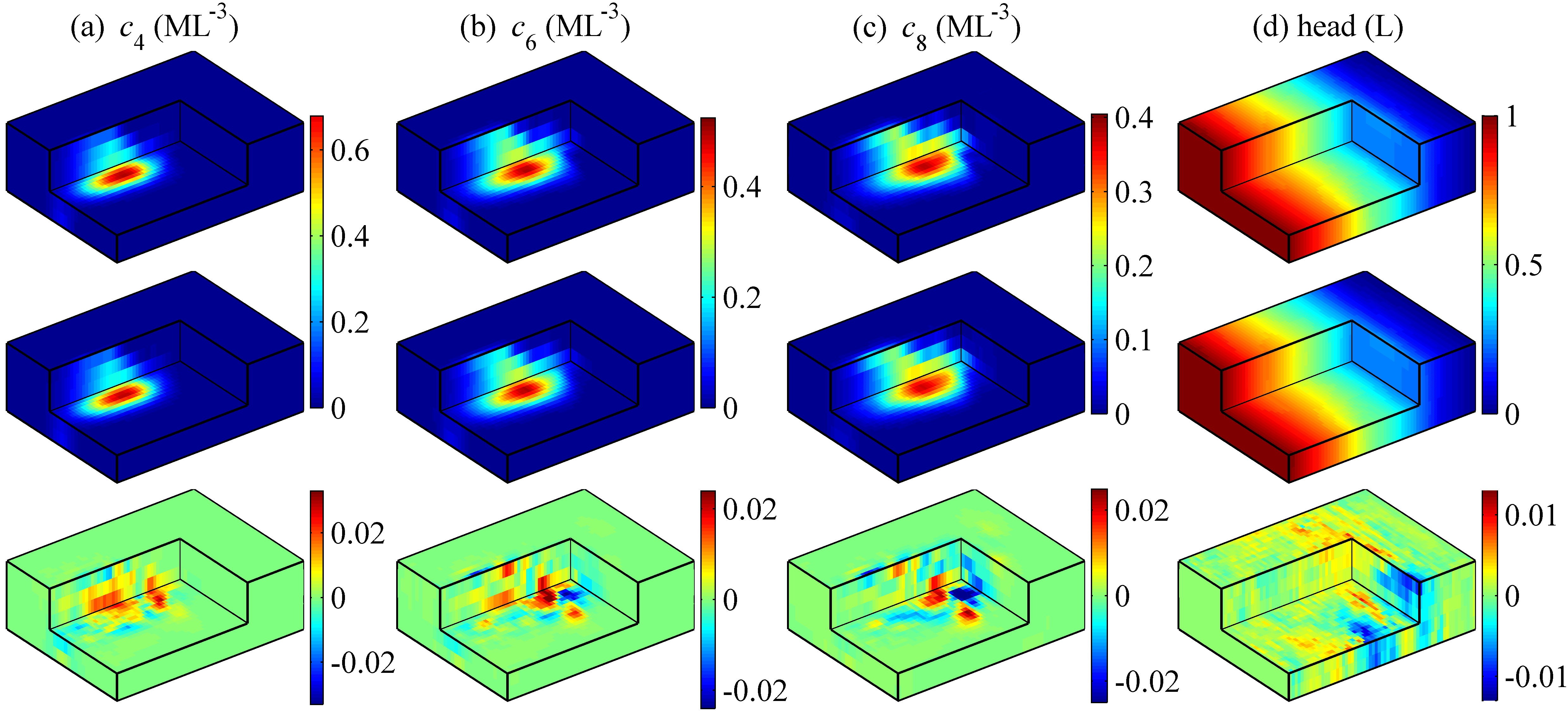}
    \caption{Comparison of the concentration $(c_t)$ fields at time $t=[4,6,8]$~(T; a-c) and the hydraulic head field (d) of a random test sample predicted by the 3-D forward model ($\mathbf{y}$) and the DRDCN network ($\hat{\mathbf{y}}$) trained using $4,000$ samples. ($\mathbf{y}-\hat{\mathbf{y}}$) denotes the network approximation error.}
    \label{fig:DRDCN-pred-3D} 
\end{figure}

The two networks' approximation accuracy for the 2-D and 3-D solute transport models is provided in Figure~\ref{fig:RMSE-R2}, 
which depicts the RMSEs, $R^2$ scores, and SSIMs evaluated on the test sets. It can be seen that DRDCN achieves lower RMSEs, higher $R^2$ 
scores, and overall higher SSIMs than DDCN when being trained using the same four training sets in both cases. For example, with $4,000$ training samples, our 
network achieves a RMSE of $0.016$, a $R^2$ score of $0.972$, and a median SSIM value of 0.994 in the 2-D case, while those obtained by the DDCN network are $0.024$, $0.933$, and 0.978, 
respectively (Figures~\ref{fig:RMSE-R2}a and~\ref{fig:RMSE-R2}c).  This implies that the DRDCN network can obtain accurate surrogates with fewer training samples 
(forward model runs) than the DDCN network. For instance, the 2-D DRDCN surrogate with $2,000$ training samples results in  
approximation accuracy comparable to that of the DDCN surrogate with $5,000$ training samples (thus with  $50$\% reduction in training samples). The saved number of training samples indicates substantial 
computational gains especially for computationally intensive forward models in subsurface modeling where one single model execution 
can take up to hours or even days. The $R^2$ scores of the DRDCN networks for the concentration fields at different time steps and the hydraulic head field are depicted in Figures~\ref{fig:RMSE-R2}e and~\ref{fig:RMSE-R2}f. It is observed that the approximation accuracy for the hydraulic head is higher than that for the concentration. This is mainly because  the hydraulic head is less sensitive to the conductivity heterogeneity compared to the concentration~\cite{Kitanidis2015}, leading to relatively smooth hydraulic fields that are easier-to-approximate.  In addition, there is no significant difference between the approximation accuracy for the concentration fields at different time steps. The DRDCN network's predictions for the output fields of 2-D and 3-D models given randomly 
selected input log-conductivity fields in the test sets are illustrated in Figures~\ref{fig:DRDCN-pred} and~\ref{fig:DRDCN-pred-3D}, 
respectively. For comparison, the predictions by the forward models and network approximation errors are also shown in each plot. 
It is observed that although the output fields are highly irregular with sharp response changes, our network is able to obtain good approximations in both cases. 

It is worth noticing that the DRDCN network achieves higher accuracy improvement compared to the DDCN network 
in the 2-D case than in the 3-D case (Figure~\ref{fig:RMSE-R2}). The low-conductivity regions are barriers for groundwater flow and solute transport, leading to irregular output fields. In comparison to the 2-D log-conductivity fields which are mainly composed of low-conductivity regions (Figure~\ref{fig:refK}c), the facies with the minimum average log-conductivity in the trimodal 3-D fields (i.e., the blue regions in Figure~\ref{fig:refK}e) is only a small proportion of the aquifer as indicated in Figure~\ref{fig:refK}g. The small proportion of low-conductivity regions in the 3-D fields thus results in relatively smoother output fields than those of the 2-D model. As a consequence, the DDCN network can also obtain a relatively accurate surrogate of the 3-D model, although its approximation error is larger than that of the DRDCN network. However, its performance greatly decreases in the 2-D model with more complex output fields. The results clearly suggest that the proposed DRDCN network performs better than the DDCN network in obtaining 
accurate surrogate results for the solute transport models with high-dimensional and highly-complex input-output relations. 
The improvement is attributed to the deeper network architecture with the help of the multilevel residual learning~\cite{Wang-etal2018} 
and residual scaling strategies~\cite{Szegedy2016}. 
With a relatively small number of training samples, the DRDCN network is able to 
provide good approximations of the 2-D and 3-D forward models. Thus, it is used together with the CAAE parameterization 
strategy in the ILUES inverse method to formulate an efficient inversion framework for estimation of a non-Gaussian conductivity field of solute transport models.

\subsection{Inversion Results}

In both 2-D and 3-D cases, the DRDCN surrogates trained with $4,000$ forward model runs are used to substitute the forward models in the CAAE-DRDCN-ILUES inversion framework. 
To assess the accuracy and computational efficiency of the surrogate-based method, the CAAE-ILUES method which evaluates the forward model rather than the surrogate model during the inversion is 
also performed.  The ensemble size of the ILUES algorithm is set to $N_e=2,000$ and $N_e=3,000$ in the 2-D and 3-D cases, respectively, to fully quantify the parametric uncertainty. 
The observations are assimilated for $N_{\rm{iter}}=30$ iterations in the 2-D case and $20$ iterations in the 3-D case. That is, the number of forward model runs required in CAAE-ILUES for the 2-D and 3-D cases are $31\times2,000=62,000$ (i.e., one prior ensemble and $N_{\rm{iter}}$ updated ensembles) 
and $21\times3,000=63,000$, respectively. The posterior log-conductivity fields are then obtained from 
the CAAE's decoder given the final latent code ensemble as input (Algorithm~\ref{algor:CAAE-DRDCN-ILUES}).

\begin{figure}[h!]
    \centering
    \includegraphics[width=0.9\textwidth]{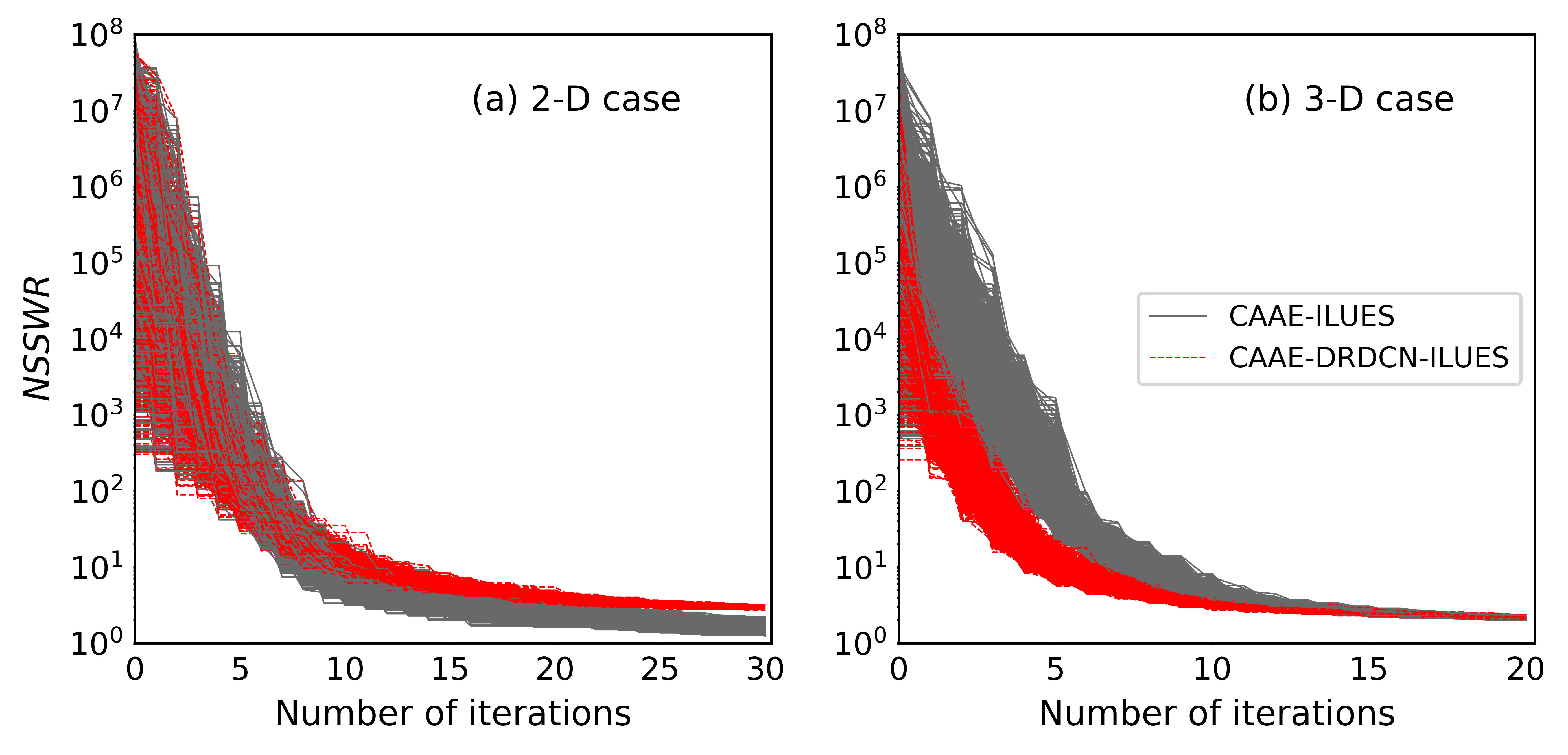}
    \caption{Convergence of the $NSSWR$ values in the ensembles of CAAE-ILUES and CAAE-DRDCN-ILUES as 
    the number of iterations increases.}
    \label{fig:SSWR} 
\end{figure}

The convergence of 
the fitting error between the model predictions and measurements as the iteration proceeds is shown 
in Figure~\ref{fig:SSWR}, where the mismatch is measured using the normalized sum of squared weighted residuals ($NSSWR$): 
\begin{linenomath*}
\begin{equation}\label{eq:sswr}
    NSSWR=\frac{1}{SSWR_{\rm{ref}}}\sum_{i=1}^{N_d}\Big(\frac{f_i(\hat{\mathbf{x}})-d_i}{\sigma_i}\Big)^2,
\end{equation}
\end{linenomath*}
where $\{d_i\}_{i=1}^{N_d}$ denote the measurements that contain measurement errors with standard deviations $\{\sigma_i\}_{i=1}^{N_d}$, and $\big\{f_i(\hat{\mathbf{x}})\big\}_{i=1}^{N_d}$ are 
the forward model predictions given the input $\hat{\mathbf{x}}$ which is generated by the decoder. Here the $SSWR$ metric is normalized using the reference value $SSWR_{\rm{ref}}$. Thus, a $NSSWR$ value approaching 1.0 suggests the convergence of the inversion process.
As discussed in sections~\ref{sec:synthetic-obs} and~\ref{sec:CAAE-parameterized}, the reference log-conductivity field has high heterogeneity versus the smoothed 
CAAE-parameterized field realizations $\hat{\mathbf{x}}$.  As a consequence, in the inversion process the $NSSWR$ values are not able to converge to the reference $NSSWR$ value (i.e., $1.0$) due to the conductivity smoothness in the CAAE-generated samples. 
 This can be seen in Figure~\ref{fig:SSWR} where 
although the $NSSWR$ values in the ensemble approximately converge in both cases, 
the converged values (the mean values are 1.68 and 2.17 in the 2-D and 3-D cases, respectively) are larger than $1.0$. Our tests showed that the $NSSWR$ values were not able to converge asymptotically to $1.0$ even when we increased the number of iterations in ILUES. The results of CAAE-DRDCN-ILUES are also shown in Figure~\ref{fig:SSWR} (the surrogate prediction $\hat{f}(\cdot)$ is used in equation~(\ref{eq:sswr})). This method converges to larger $NSSWR$ values (the mean values are $2.95$ and $2.19$ in the 2-D and 3-D cases, respectively) mainly due to the approximation errors of the DRDCN surrogate model. Accounting for the approximation errors in the inversion is expected to further improve the estimation accuracy~\cite{Cui2011,zhang2016}. This can be realized by building  a  model for the approximation errors and then accounting for the approximation error contribution to the covariance matrix $\mathbf{C}_{\text{D}}$~\cite{Cui2011}. Alternatively, one can employ a Bayesian training strategy (e.g., Stein
variational gradient descent~\cite{LiuSVGD2016}) for the DRDCN network as in~\citeA{ZHU2018415}, which can provide multiple sets of tuned network parameters and thus an estimation of the variance in the prediction. The uncertainty estimate can subsequently be incorporated in $\mathbf{C}_{\text{D}}$ as in~\citeA{zhang2016} to alleviate the influence of approximation errors. The training of Bayesian networks, however, is more expensive than that of non-Bayesian networks~\cite{ZHU2018415}.

\begin{figure}[h!]
    \centering
    \includegraphics[width=\textwidth]{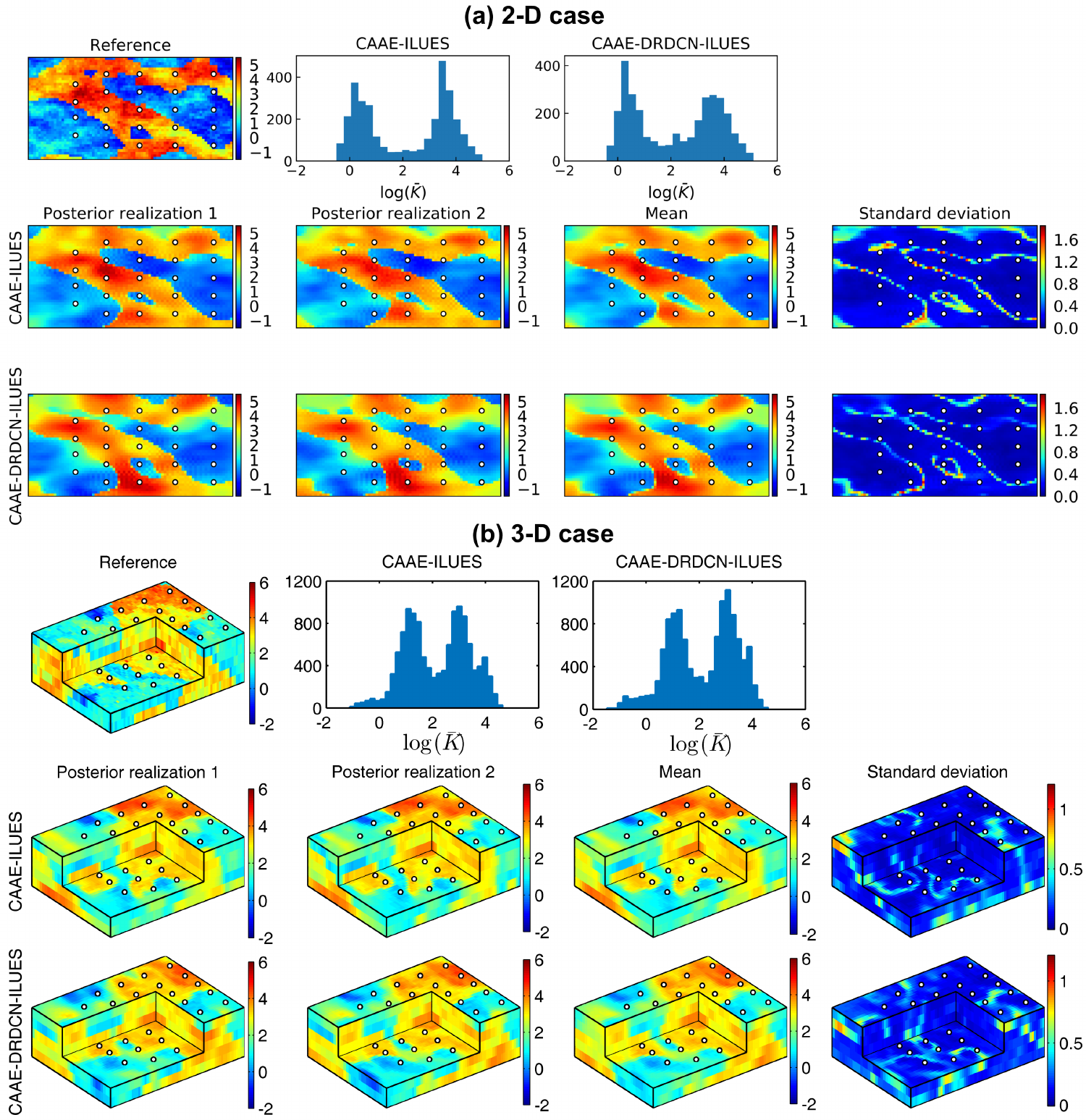}
    \caption{Two posterior log-conductivity realizations, ensemble mean, and ensemble standard deviation 
    obtained from the CAAE-ILUES and the CAAE-DRDCN-ILUES methods. The histograms are for the mean log-conductivity values ($\log(\bar{K})$) of all posterior realizations. The reference fields shown in (a) and (b) are those
    in Figures~\ref{fig:refK}a and~\ref{fig:refK}d, respectively. The circles denote the projections of the output measurement locations on the horizontal plane.}
    \label{fig:Kest} 
\end{figure}

\begin{figure}[h!]
    \centering
    \includegraphics[width=\textwidth]{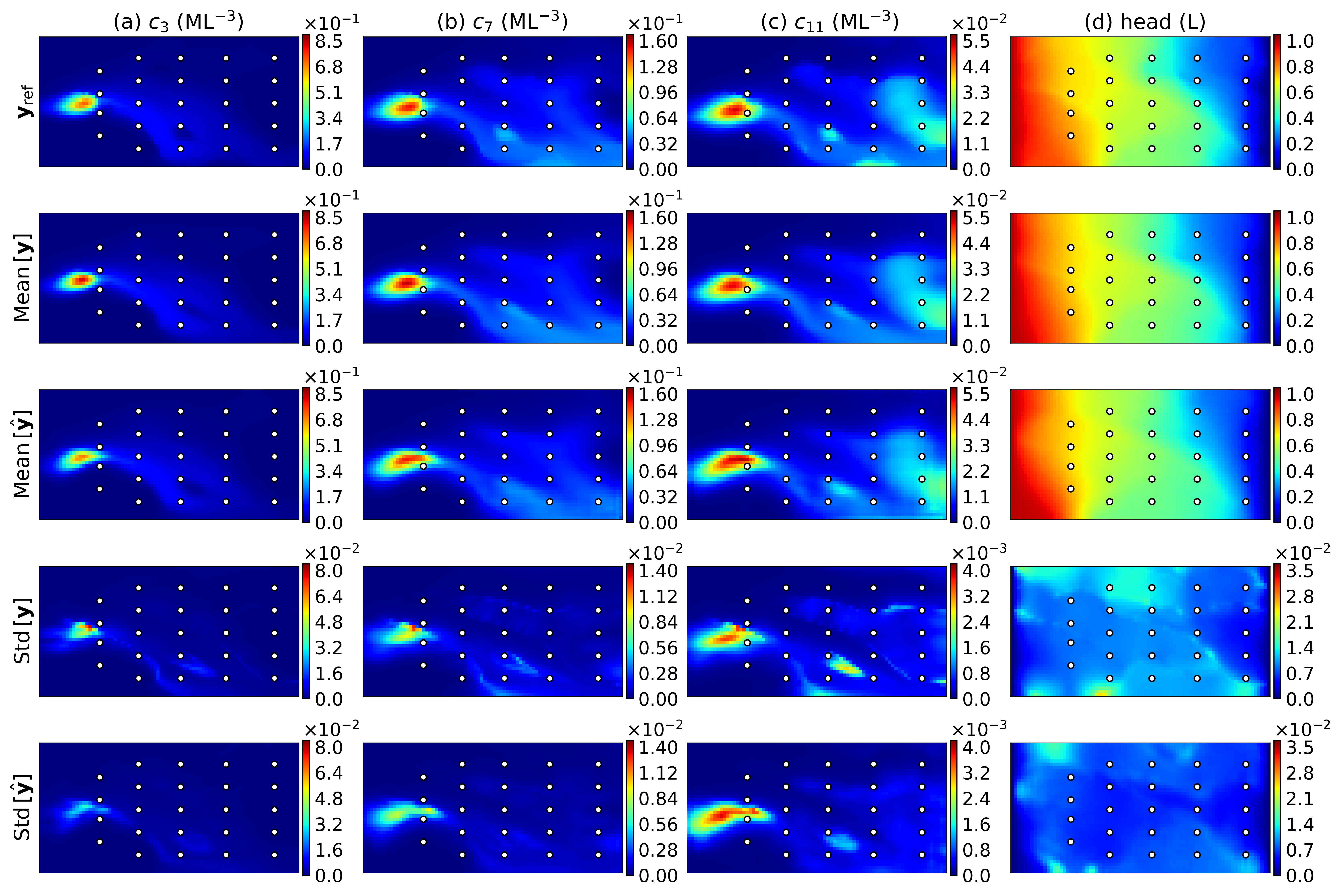}
    \caption{Ensemble mean and standard deviation (Std) of the concentration $(c_t)$ fields at time $t=[3,7,11]$~(T; a-c) and 
    hydraulic head field (d) of the 2-D model obtained from CAAE-ILUES (Mean[$\mathbf{y}$], Std[$\mathbf{y}$]) and CAAE-DRDCN-ILUES (Mean[$\hat{\mathbf{y}}$], Std[$\hat{\mathbf{y}}$]). $\mathbf{y}_{\rm{ref}}$ denotes the output fields of the reference model with the log-conductivity field shown in Figure~\ref{fig:refK}a. The circles denote the measurement locations.}
    \label{fig:mean-std} 
\end{figure}

Two posterior log-conductivity realizations, the ensemble mean, ensemble standard deviation, and histogram of the log-conductivities in the ensemble mean field of the CAAE-ILUES method for the two cases
are illustrated in Figure~\ref{fig:Kest}. The reference fields and the locations of the output 
measurements collected to infer the conductivity field are also shown in the plot to facilitate the comparison. 
It can be seen that in both cases the CAAE-ILUES can successfully capture the spatial distribution of the high-conductivity and low-conductivity regions as well as
the conductivities within these regions. This is also indicated by the multimodal distribution of the log-conductivities in the mean fields. Due to the relatively sparse measurements 
(i.e., only $24$ observation wells are placed in the domain with thousands of gridblocks), the local conductivity 
heterogeneity, the location of the boundaries between high-conductivity and low-conductivity regions, the mean values of the modes in the histograms may not be accurately retrieved. 
As expected, the conductivity estimation in regions where no information is collected is less accurate with a larger estimation uncertainty 
than the estimation near the the output measurement locations. 
The results imply that the CAAE-ILUES algorithm performs well for this inverse problem but needs a 
large number of forward model evaluations. We show next the results 
of CAAE-DRDCN-ILUES in which the DRDCN surrogate models are used to substitute 
the 2-D and 3-D forward models. 
Various statistics for the same reference fields as in CAAE-ILUES 
are depicted in the third row of Figures~\ref{fig:Kest}a (2-D case) and~\ref{fig:Kest}b (3-D case).
Similarly, it can be seen that in both cases the surrogate-based 
framework successfully captured the multimodal features and identified the high-conductivity and low-conductivity regions as well as the conductivities within these regions.

\begin{figure}[h!]
    \centering
    \includegraphics[width=\textwidth]{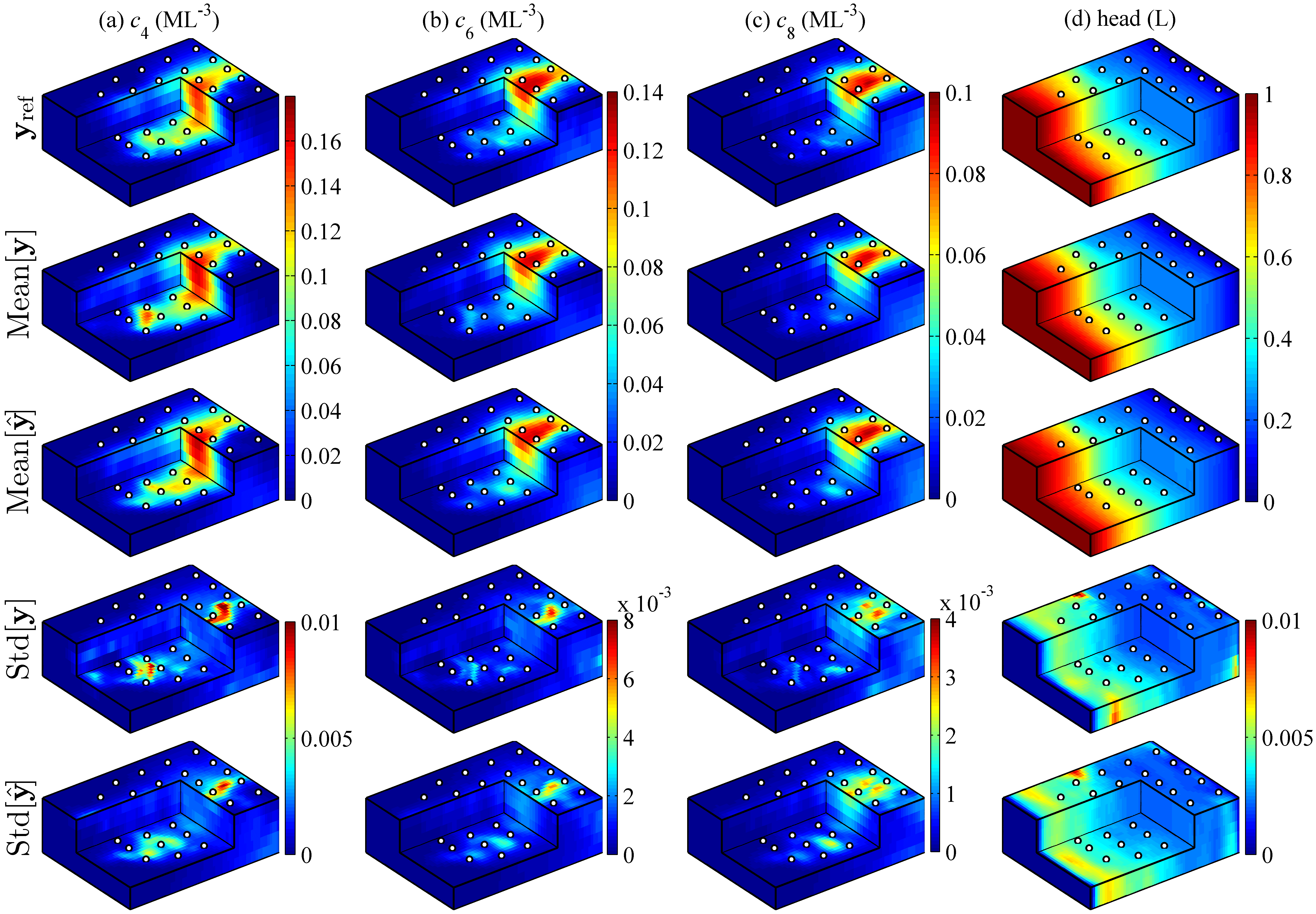}
    \caption{Ensemble mean and standard deviation (Std) of the concentration $(c_t)$ fields at time $t=[4,6,8]$~(T; a-c) and 
    hydraulic head field (d) of the 3-D model obtained from CAAE-ILUES (Mean[$\mathbf{y}$], Std[$\mathbf{y}$]) and 
    CAAE-DRDCN-ILUES (Mean[$\hat{\mathbf{y}}$], Std[$\hat{\mathbf{y}}$]). $\mathbf{y}_{\rm{ref}}$ denotes the output 
    fields of the reference model with the log-conductivity field shown in Figure~\ref{fig:refK}d. The circles denote the projections of the measurement locations on the horizontal plane.}
    \label{fig:mean-std-3D} 
\end{figure}

Figures~\ref{fig:mean-std} and~\ref{fig:mean-std-3D} show the output ensemble mean and standard deviation estimates of the CAAE-ILUES and 
CAAE-DRDCN-ILUES methods for the 2-D and 3-D models, respectively.  
The reference output fields and the observation locations are also shown to facilitate the analysis of results. Notice that the statistics 
of the surrogate-based method are computed using the outputs predicted by the surrogate model rather than the forward model. 
It can be observed that the CAAE-DRDCN-ILUES method achieves similar ensemble mean estimates to those of the CAAE-ILUES method 
which successfully reproduce the main patterns of the reference output fields in the two cases.
Similar to the estimation of the conductivity field, a higher reproduction accuracy and a lower estimation uncertainty are observed near the observation wells than those in regions where no information is collected.  

The results indicate that the CAAE network is able to reconstruct well  non-Gaussian 
conductivity fields with different heterogeneity patterns and therefore the CAAE-ILUES method can obtain good inversion results in the 2-D and 3-D cases. However, for the two high-dimensional and highly-nonlinear 
inverse problems considered here, more than $60,000$ forward model runs are needed in each case, leading to a high computational cost.  
On the contrary, the surrogate-based framework CAAE-DRDCN-ILUES can provide similar inversion results with much less computation costs. To assess the computational efficiency of the surrogate method, let $N_{\text{C-I}}$ and $N_{\text{C-D-I}}$ denote the number of forward model runs needed by CAAE-ILUES and CAAE-DRDCN-ILUES, respectively. Also, the $C_{\text{f}}$ and $C_{\text{train}}$ represent the computational cost of one forward model run and of training the surrogate model, respectively. The computation time of the original ($C_{\text{C-I}}$) and surrogate-based ($C_{\text{C-D-I}}$) ILUESs are written as 
\begin{linenomath*}
\begin{equation}\label{eq:C-ILUES}
    C_{\text{C-I}}=N_{\text{C-I}}C_{\text{f}},
\end{equation}
\end{linenomath*}
and
\begin{linenomath*}
\begin{equation}\label{eq:C-sur-ILUES}
    C_{\text{C-D-I}}=N_{\text{C-D-I}}C_{\text{f}}+C_{\text{train}},
\end{equation}
\end{linenomath*}
respectively (the training time of CAAE parameterization model is the same for the two methods, thus it is not included in the equations). Thus, $C_{\text{C-I}}=1.74\times10^5$~s and $C_{\text{C-D-I}}=1.55\times10^4$~s in the 2-D case ($N_{\text{C-I}}=62,000$, $N_{\text{C-D-I}}=4,000$, $C_{\text{f}}=2.8$~s, and $C_{\text{train}}=4.32\times10^3$~s), and $C_{\text{C-I}}=8.25\times10^5$~s and $C_{\text{C-D-I}}=9.02\times10^4$~s in the 3-D case ($N_{\text{C-I}}=63,000$, $N_{\text{C-D-I}}=4,000$, $C_{\text{f}}=13.1$~s, and $C_{\text{train}}=3.78\times10^4$~s). The computational savings obtained by the surrogate-based method ($\frac{C_{\text{C-I}}-C_{\text{C-D-I}}}{C_{\text{C-I}}}\times100\%$) in the 2-D and 3-D cases are both about $90\%$. When applying the method to computationally more intensive forward models (i.e., when $C_\text{f}$ is large), more savings of computation time 
\begin{linenomath*}
\begin{equation}\label{eq:C-saving}
    C_{\text{saving}}=C_{\text{C-I}}-C_{\text{C-D-I}}=(N_{\text{C-I}}-N_{\text{C-D-I}})C_{\text{f}}-C_{\text{train}},
\end{equation}
\end{linenomath*}
can be expected as $N_{\text{C-I}}\gg N_{\text{C-D-I}}$.

\section{Conclusions}\label{sec:conclusions}
In this study, we propose an integrated inversion framework for efficient characterization of solute transport in non-Gaussian conductivity fields with multimodal distributions. 
In the proposed framework, a CAAE network is developed for parameterization of the conducitivity field 
using a low-dimensional latent representation. In addition, a DRDCN network is developed for surrogate modeling 
of the solute transport model with high-dimensional and highly-complex input-output mappings. The two networks 
are combined with the ILUES inversion method~\cite{zhang2018ILUES} to formulate the CAAE-DRDCN-ILUES 
inversion framework. To improve the networks' performance for approximating the highly-complex mappings in the 
problems considered, in both network architectures, 
we adopt a multilevel residual learning structure referred to as residual-in-residual dense 
block~\cite{Wang-etal2018}. 
This structure can effectively ease the training process allowing for a large network depth which has the potential to substantially increase the network's capability to approximate highly-complex mappings. 

The performance of the proposed method is evaluated using  2-D and 3-D solute transport modeling with
non-Gaussian conducitivity fields that have different patterns of conductivity heterogeneity. The results indicate that the CAAE is capable of representing a non-Gaussian conductivity 
field using a low-dimensional latent vector, though this is at the cost of the smoothness of local heterogeneity. The DRDCN network shows a superior performance over the 
DDCN network proposed in our previous studies~\cite{mo2019inverse,mo2019UQ,ZHU2018415,zhu2019physics} in obtaining accurate surrogate models.
The residual-in-residual dense block structure greatly improves the network's capacity in 
approximating highly-complex mappings. The application of the CAAE-DRDCN-ILUES method for estimation of 
the 2-D and 3-D non-Gaussian conductivity fields shows that it can obtain similar 
inversion results and predictive uncertainty estimations 
to those obtained by the original inverse method without surrogate modeling. The integrated 
method is highly efficient since the training of the surrogate model requires only a small number of forward model runs. 
The solute transport models considered in this work are relatively fast in order to quickly test the proposed method in a reasonable time. When applying 
to computationally more intensive forward models, significant computational gains can be expected. 

Note that incorporating the surrogate model in the inversion introduces an additional source of uncertainty due to approximation errors. Consideration of these approximation errors in the inversion process can be beneficial in improving the accuracy and reliability of the results~\cite{Cui2011,zhang2016}. To this end, even if not considered or demonstrated herein, one can build a  model for the approximation errors~\cite{Cui2011} or employ a Bayesian version of the DRDCN surrogate~\cite{ZHU2018415,zhu2019physics} for estimating the variance in the prediction, so as to incorporate the approximation errors in the inversion formula.
In the current work, the presented methods were demonstrated using synthetic problems. Their potential use in practical applications and other complex systems beyond groundwater solute transport deserves further exploration due to their data-driven and nonintrusive nature.

\acknowledgments
S.M. acknowledges partial financial support from the Center for 
Informatics and Computational Science (CICS) at the University of Notre Dame where this work was performed. The work of N.Z. was supported 
from the Defense Advanced Research Projects Agency under the Physics of Artificial Intelligence 
program (contract HR$00111890034$). The computing was supported via an AFOSR-DURIP award with additional   resources   provided by the University of Notre Dame Center for 
Research Computing. The work of X.S. and J.W. was supported by the 
National Natural Science Foundation of China (No. $41730856$ and $41672229$). The authors would like to thank the three anonymous reviewers and the Editors for their helpful comments.
The codes and data used in this work are available at \url{https://github.com/cics-nd/CAAE-DRDCN-inverse}.

\appendix

\section{Deep dense convolutional network}\label{appendix:DDCN}
The deep dense convolutional network (DDCN) for surrogate modeling~\cite{mo2019inverse,mo2019UQ,ZHU2018415,zhu2019physics} 
is based on a dense connection structure called dense block~\cite{Huang2016}. The dense block 
introduces connections between its internal nonadjacent layers to enhance the information propagation 
through the network, so as to reduce the training sample size 
required to obtain desired approximation accuracy~\cite{Huang2016}. An illustration of the dense block 
structure is shown in Figure~\ref{fig:net4surrogate}a. The difference between the dense blocks used 
in the DDCN network and in the DRDCN network proposed in this study is that, in DDCN, the input 
feature maps of the dense block's last layer are concatenated to its output feature maps to be 
fed into the next layer; while in DRDCN only the output feature maps of the dense block's last layer are 
passed to the next layer to allow an element-wise addition operation 
in the residual learning strategy (see section~\ref{sec:RRDB}). 

We adopt the DDCN network architecture employed in our previous study~\cite{mo2019inverse} 
which was used for surrogate modeling of a 2-D solute transport model with Gaussian conductivity fields. The ReLU instead of Mish activation was used in DDCN.
The network is composed of $27$ convolutional layers with three dense blocks. For the 3-D case considered in this study, we directly replace the 2-D convolutional layers in the network with the 3-D convolutional layers. 
More details about the network architecture can be found in~\citeA{mo2019inverse}. When training the DDCN network, we use the same settings as in the DRDCN network. That is, the network is trained using the $L_1$ loss function defined in equation~(\ref{eq:l1-loss}) for $200$ epochs  in the 2-D case and $300$ epochs in the 3-D case  with the Adam optimizer~\cite{kingma2014-adam}. The batch size is $32$ and the initial learning rate is $5\times 10^{-3}$. A learning rate scheduler which drops ten times on plateau during training is used.  
Python codes of DDCN are available at \url{https://github.com/cics-nd/cnn-inversion}.

\bibliography{my}

\end{document}